\documentclass[letterpaper]{JHEP3}

\let\ifpdf\relax
\usepackage{ifpdf}

\usepackage{graphicx}
\usepackage{amsmath}

\newcommand{\url}[1]{\href{#1}{#1}}

\def\simgt{\mathrel{\lower2.5pt\vbox{\lineskip=0pt\baselineskip=0pt
           \hbox{$>$}\hbox{$\sim$}}}}
\def\simlt{\mathrel{\lower2.5pt\vbox{\lineskip=0pt\baselineskip=0pt
           \hbox{$<$}\hbox{$\sim$}}}}
           
\title{A Stealth Supersymmetry Sampler}

\author{}
\author{JiJi Fan\\
        Department of Physics, Princeton University, Princeton, NJ, 08540\\
        E-mail: \email{jijifan@princeton.edu}}
\author{Matthew Reece\\
Department of Physics, Harvard University, Cambridge, MA, 02138\\
 E-mail: \email{mreece@physics.harvard.edu}}
 \author{Joshua T. Ruderman\\
 Berkeley Center for Theoretical Physics, Department of Physics,\\
 and Theoretical Physics Group, Lawrence Berkeley National Laboratory,\\
 University of California, Berkeley, CA 94720\\
 E-mail: \email{ruderman@berkeley.edu}}

\abstract{The LHC has strongly constrained models of supersymmetry with traditional missing energy signatures. We present a variety of models that realize the concept of Stealth Supersymmetry, i.e. models with $R$-parity  in which one or more nearly-supersymmetric particles (a ``stealth sector") lead to collider signatures with only a small amount of missing energy. The simplest realization involves low-scale supersymmetry breaking, with an $R$-odd particle decaying to its superpartner and a soft gravitino. We clarify the stealth mechanism and its differences from compressed supersymmetry and explain the requirements for stealth models with high-scale supersymmetry breaking, in which the soft invisible particle is not a gravitino. We also discuss new and distinctive classes of stealth models that couple through a baryon portal or $Z'$ gauge interactions. Finally, we present updated limits on stealth supersymmetry in light of current LHC searches. }

\newcommand{\beq}{\begin{equation}}
\newcommand{\eeq}{\end{equation}}
\newcommand{\beqs}{\begin{eqnarray}}
\newcommand{\eeqs}{\end{eqnarray}}
\newcommand{\gsim}{\stackrel{>}{_\sim}}
\newcommand{\lsim}{\stackrel{<}{_\sim}}

\newcommand{\bary}{\bar{Y}}

\newcommand {\unit} [1] {\; \mathrm {#1}}

\newcommand{\met}{\mbox{${\rm \not\! E}_{\rm T}$}}

\begin{document}

\section{Introduction}
\label{sec:intro}
\setcounter{equation}{0}
\setcounter{footnote}{0}

Supersymmetry (SUSY) is one of the most compelling possible explanations for the stability of the hierarchy between the weak scale and Planck scale, and a leading contender for a new principle of nature that could be discovered at the Large Hadron Collider (LHC). The LHC is operating successfully, with over 5/fb of data recorded per experiment and many analyses of 1/fb of data so far, which have substantially altered the viable parameter space for many models of new physics. In the case of SUSY, already in early summer 2011 ATLAS set limits of above 1 TeV on gluino and (light generation) squark masses using only 165/pb of data~\cite{ATLAS165}. This 1 TeV exclusion marked a major milestone on the path to exploring new physics at the weak scale, and a plethora of new results followed. Recent theoretical analyses~\cite{Essig:2011qg,Kats:2011qh,Brust:2011tb,Papucci:2011wy,Bi:2011ha,Desai:2011th} have helped to map out the excluded region of parameter space, and have shown that (depending on assumptions about which states are degenerate and how they decay) light stops and sbottoms may already be excluded up to masses of about 300 GeV\@. Because the stop cancels the largest divergence in the Higgs mass, this indicates that the LHC is poised to begin making definite statements about naturalness in standard supersymmetric scenarios.

As emphasized in~\cite{Kats:2011qh,Papucci:2011wy}, two types of studies are currently setting very strong limits on supersymmetric models. The first are searches for jets and missing transverse momentum~\cite{Aad:2011ib,CMS-PAS-SUS-11-004,CMS-PAS-SUS-11-003}, which suppress standard model (SM) backgrounds by requiring {\em very} large missing transverse momentum and several very hard jets. These analyses demonstrate the raw power the LHC achieves simply by operating at energies never probed by any previous collider. The second type of study that sets very strong limits focuses on clean signals that have almost no SM background. The exemplar is a search for same-sign dileptons~\cite{CMS-PAS-SUS-11-010,arXiv:1110.6189}, which is a very interesting probe of final states with multiple top quarks (among others); for a more specialized class of models, diphoton searches~\cite{CMS-PAS-SUS-11-009,ATL-PHYS-SLIDE-2011-523} play a similar role. The combination of jets and missing $p_T$ with same-sign dileptons is already enough to exclude large parts of the parameter space of natural SUSY, assuming $R$-parity and decays to invisible particles that escape the detector. 

The present stringent LHC limits leave three options for SUSY theories. The first is that SUSY is natural, with stops canceling the largest loop corrections to the Higgs mass, and $R$-parity and traditional collider signatures are present. To evade bounds, this requires {\em flavored} SUSY breaking, with squarks of the first two generations much heavier than those of the third. The second option is that SUSY is less natural, and all squarks including the stops are heavy, with associated fine-tuning in the Higgs mass.\footnote{In some scenarios, a large higgs quartic coupling could be generated, which ameliorates the fine tuning.  This can be consistent with the possible higgs discovery near $m_h \sim 125$~GeV~\cite{Hall:2011aa}.} The third case is that SUSY is natural, and the squarks of the first and second generation remain light, but the collider signatures are altered so that the current searches are evaded: in particular, the classic missing energy signature does not apply. One traditional example of the third class is $R$-parity violation (RPV). However, as we recently suggested, the third possibility could also be realized in SUSY theories with $R$-parity, using the mechanism of {\it Stealth SUSY}~\cite{Stealth}. Stealth SUSY opens up the possibility of natural SUSY with light stops and a simple flavor structure. In this paper, our goal is to look at a broader class of stealth models. Originally, we considered only models with one singlet chiral superfield $S$ as the stealth sector, interacting with the MSSM via superpotential couplings to either Higgses or vectorlike matter, and decaying to a gravitino. Here, after reviewing the stealth mechanism, we will present several variations. We will drop the assumption of low-scale SUSY breaking and explain how high-scale SUSY breaking, especially anomaly mediation~\cite{Randall:1998uk,GLMR}, can also lead to stealth phenomenology. We will also consider models in which the stealth sector contains fields with baryon number, which lead to events with extremely high jet multiplicities. Furthermore, we will discuss some models with new gauge interactions, either a $Z'$ or a confining sector that dynamically generates massive singlets. Finally, we will also present an updated discussion of the collider physics of stealth SUSY in light of current LHC searches.

Before ending the introduction, we want to comment on the implication for stealth SUSY of the recently observed possible Higgs bump at 125 GeV~\cite{CMS-PAS-HIG-11-032, ATLAS-CONF-2011-163}. Stealth SUSY itself could be completely natural with a light stop and small radiative correction to the Higgs mass. To raise the Higgs mass above the $Z$ boson mass, additional tree-level contributions to the Higgs mass need to be incorporated into the stealth scenario. In this paper, we will not discuss how to raise the Higgs mass to 125 GeV, as it could be easily achieved by adding an additional module to the stealth SUSY framework. Interestingly, one stealth SUSY model which could accommodate a 125 GeV Higgs has already been constructed in~\cite{Csaki:2012fh}, realizing the ``stealth stop" scenario.

\FIGURE[h]{
\includegraphics[scale=0.75]{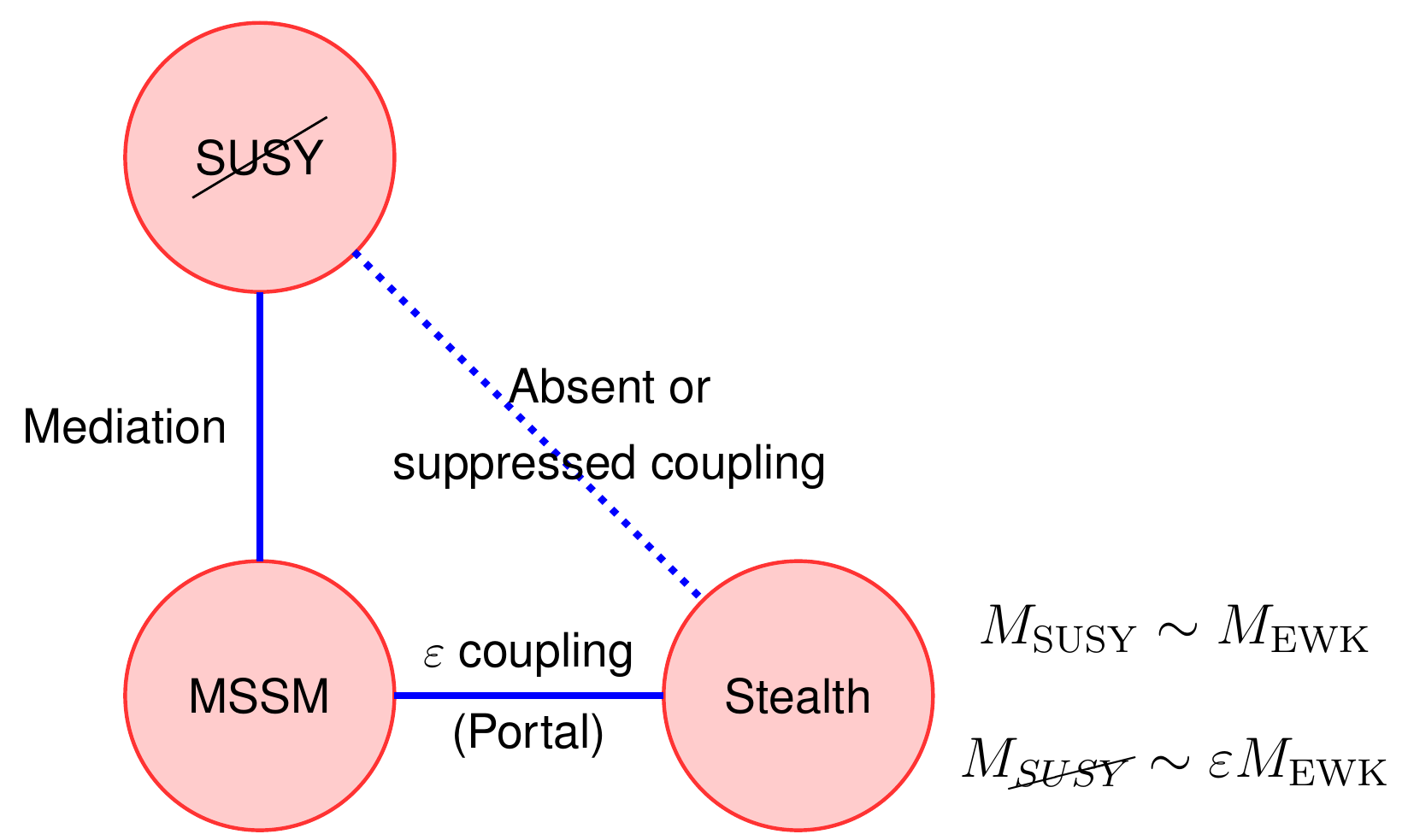}
\caption{A schematic of the sectors involved in a general stealth model. Flavor-blind mediation gives rise to standard MSSM soft SUSY-breaking terms, but the soft terms in the stealth sector are suppressed relative to this. The MSSM and the stealth sector are weakly coupled, and the size of soft terms in the stealth sector is suppressed relative to the supersymmetric mass scale of the stealth sector by a weak-coupling factor.}
\label{fig:genericstealth}
}

\section{General Features of Stealth SUSY}

\subsection{The Stealth Mechanism}

The basic ingredients of stealth SUSY are illustrated in Figure~\ref{fig:genericstealth}. The key requirement is a set of particles that are nearly degenerate with their superpartners (with supersymmetric masses much larger than SUSY-breaking splittings). We will refer to the complete set of such fields that feel only small SUSY-breaking as the ``stealth sector." It may be as simple as a single chiral superfield, as in examples discussed in Ref.~\cite{Stealth}, or it could be a rich sector with one or more gauge groups and many matter fields. In any case, there must be a portal through which the lightest ($R$-odd) MSSM superpartner (lighest ordinary superpartner or LOSP) can decay to a particle in the stealth sector. After this, a decay chain within the stealth sector can occur, but it must end with a massive $R$-odd stealth particle decaying to a nearly degenerate $R$-even state plus a light $R$-odd state. In the simplest realization, this final $R$-odd state is the gravitino, but we will be interested in more general models. Finally, $R$-even stealth states produced in the decay chain must in turn be able to decay back to SM fields. The outcome should be that missing energy is carried away only by the light $R$-odd particle terminating the decay chain, which has momentum suppressed by the small phase space available in the decay that produced it. (General decay chains in a complex enough stealth sector could involve multiple such lightest $R$-odd particles escaping the chain; as long as the splittings are sufficiently small and the typical multiplicity is low, SUSY can still be hidden at colliders.)

\subsection{Stealth SUSY Is Not Compressed SUSY}

It is well-known that, for standard gravity-mediated MSSM spectra, collider signals are more difficult to observe as the masses are compressed. For instance, a gluino decaying to a bino and two quarks, ${\tilde g} \to q{\bar q}{\tilde B}$, is most constrained if the bino is nearly massless, in which case a significant fraction of the gluino's energy goes into invisible momentum from the bino. As the mass splitting is reduced, the typical missing energy in the event is reduced, and limits from LHC searches grow weaker. Recent discussions of limits on compressed scenarios can be found in~\cite{LeCompteMartin}. Superficially, stealth SUSY might sound like a special case of compressed SUSY: mass splittings are small, missing $E_T$ is reduced, and limits are weaker. However, there is a crucial kinematic difference, associated with the fact that in standard compressed SUSY, the invisible particle is a {\em heavy} decay product, whereas in stealth SUSY the invisible particle is very light. This ensures that the reduced missing $E_T$ of stealth SUSY is much more robust against effects like initial state radiation.

To clarify this difference, we will review some basic relativistic kinematics and rules-of-thumb for hadron collider physics. First, consider the decay of a heavy particle of mass $M$ to a particle of mass $m = M - \delta M$ and a massless particle. In the rest frame, the momentum of the daughter particles is
\beq
p = \frac{M^2 - m^2}{2 M} = \delta M - \frac{\delta M^2}{2 M}.
\eeq
Thus, in the limit of a small mass splitting, the momentum of the daughters is approximately equal to the mass splitting. Of course, the {\em energy} of the heavy daughter is then approximately equal to its mass, which is of order $M$, so the heavy daughter moves slowly, with velocity $\delta M/M$. The result is that when boosting to the lab frame (with Lorentz factor $\gamma$), the momentum of the heavy daughter is approximately equal to that of the parent particle, whereas the massless daughter will have momentum of order $\gamma~\delta M$, since all of its momentum components in the parent rest frame were of order $\delta M$.

What are the typical (transverse) boosts involved in going to the lab frame for SUSY cascade decays? Even without running a simulation or numerically integrating over PDFs, we can give a semi-analytic answer to this question. Consider, as an example, production of gluino pairs at a proton--proton collider. There are two competing effects that determine the typical momentum of a gluino. The first is phase space, which grows with increasing momentum; the second are parton luminosities, which determine how often hard enough quarks or gluons can be extracted from the protons, and fall rapidly at large momentum. Locally, we can approximate that the parton luminosities fall like power laws with exponents that will typically be between -3 and -6, depending on details of which partons are colliding and in what energy range~\cite{CampbellHustonStirling}. In fact, one obtains a reasonably good model of the shape of the produced gluino $p_T$ distribution, $d\sigma/dp_T$, simply by multiplying a factor for the growth of phase space above threshold ($\propto p_T$) by a power-law falloff arising from integrating over parton distributions ($\propto (p_T^2 + m^2)^{-k}$), which amounts to assuming a constant matrix element in addition to a simple power-law for the parton luminosities. We give an example in Figure~\ref{fig:compressed}. A more detailed explanation of why this simple ansatz for the $p_T$ distribution is so numerically accurate can be found in~\cite{Marmoset}. The upshot, however, is that the average $p_T$ of a produced gluon can be found by maximizing the function $x(1 + x^2)^{-k}$, where $x = p_T/m$; numerically, $k = 6$ works fairly well and leads to a typical $p_T \approx 0.3 m$. (The distribution is skewed to the right, so the {\em mean} $p_T$ is somewhat, but not dramatically, larger.)

This means that in the case of {\em compressed} SUSY, the invisible particles are not really soft at all in the lab frame! For a 600 GeV gluino, for instance, the typical $p_T$ of a bino arising from the gluino decay is 200 GeV, which would be a rather large amount of missing energy. As explained clearly in~\cite{Alwall:2008ve,Alwall:2008va}, the reason why limits get weaker here is that when gluinos are pair-produced, their transverse momenta balance, so that in the compressed case their bino daughters also have nearly equal transverse momenta and the net missing $E_T$ is small, even though the individual escaping particles are hard. This reduced missing $E_T$ is not robust, as in the presence of ISR jets, the gluino $p_T$s are no longer equal, and sizable missing $E_T$ can be produced. Thus, compressed SUSY weakens the limits from standard searches, but still leads to many events with large missing $E_T$ that can be searched for.

\FIGURE[h]{
\includegraphics[width=\textwidth]{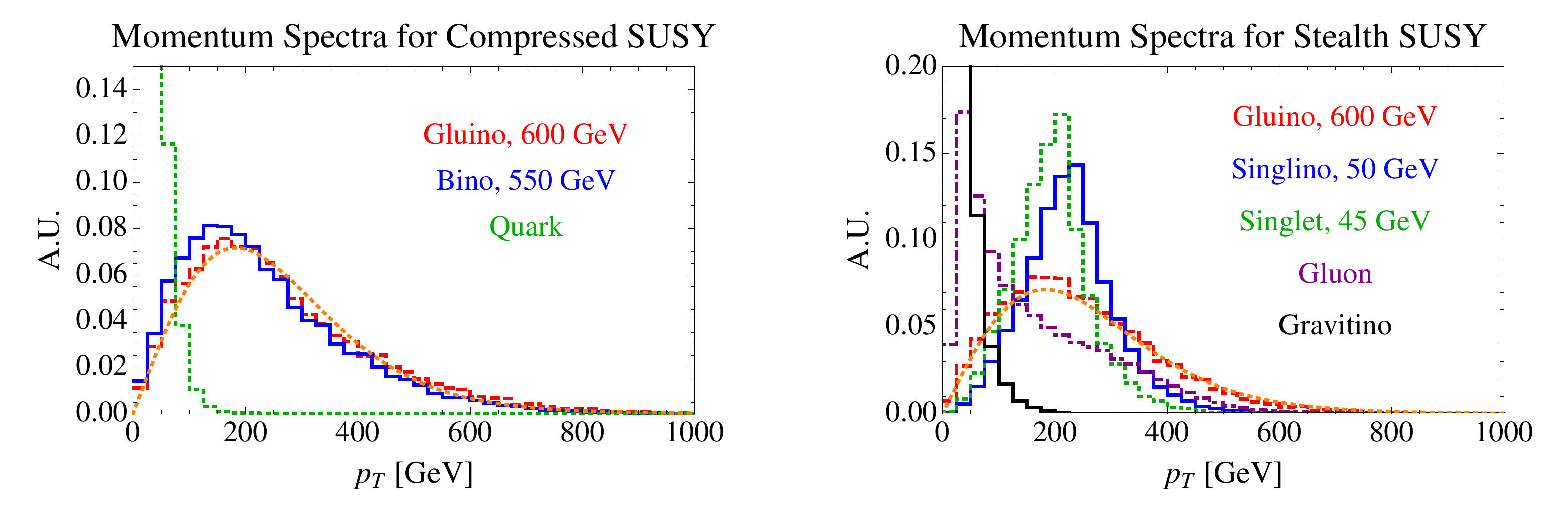}
\caption{Momentum spectra in compressed theories. At left: standard compressed SUSY, with nearly degenerate gluino and bino and the decay chain $\tilde g \to q {\bar q} {\tilde B}$. The bino momentum is typically very close to that of the gluino, and is not soft. The orange dotted curve is a simple ansatz $d\sigma/dp_T \propto p_T(p_T^2 + m^2)^{-6}$ to illustrate the characteristic interplay of phase space and steeply-falling parton luminosities. At right: stealth SUSY, with the same gluino mass, now decaying in the chain $\tilde g \to g \tilde S$, $\tilde S \to S \tilde G$, and $S \to gg$. Note that the gravitino, the invisible particle in the stealth case, has a $p_T$ distribution resembling that of a quark in the usual compressed SUSY case, and is very soft.}
\label{fig:compressed}
}

In the case of {\em stealth} SUSY, on the other hand, the invisible particles are soft in the lab frame, and the suppression of missing $E_T$ results not from a cancellation but from the complete absence of high-momentum invisible particles in the event. In particular, because the typical transverse boost of the original parent particle (gluino, for instance) is not large, we can estimate the boost of the stealth parent (singlino $\tilde S$, in the models of~\cite{Stealth}) to be $\gamma \sim m_{\tilde g}/m_{\tilde S}$. Then the lab-frame momentum of the invisible particle is
\beq
p_{\rm invis} \sim \gamma~\delta M \sim m_{\tilde g} \frac{m_{\tilde S} - m_S}{m_{\tilde S}}.
\eeq
Compared to the bino momentum in the compressed case, which was $\sim 0.3~m_{\tilde g}$, this can be made arbitrarily small by taking the stealth splitting small. The reduced missing $E_T$ in the stealth case is much more robust, as it is independent of any amount of radiation or the structure of the cascade decay. We illustrate some of the relevant $p_T$ spectra in Figure~\ref{fig:compressed}.

\subsection{Stealthy SUSY Breaking}
\label{subsec:stealthbreaking}

Having argued that the stealth mechanism is robust from the standpoint of suppressing missing energy, the next general issue is whether it is robust from a model-building point of view. The setting in which stealthy physics arises with the least effort is low-scale SUSY breaking, which always has a light gravitino that appears in the decay of a particle to its superpartner. Furthermore, the low scale of SUSY breaking can explain why dangerously large soft terms in the stealth sector are absent. One still has to explain the supersymmetric masses in the stealth sector, which are near the electroweak scale either by accident or through common underlying physics. The simplest explanation is to generate them in the same way one generates the MSSM $\mu$-term; however, to preserve stealthy physics, one would then need to require that $B\mu$ be small. Ordinarily, in gauge mediated model-building, one declares the $\mu/B\mu$ problem to be solved in models that naturally generate $B\mu \sim \mu^2$, so we need a more stringent criterion. Nonetheless, palatable solutions exist with $B\mu \ll \mu^2$, including retrofitting~\cite{retrofitting}, messenger models with $R$-symmetries~\cite{KomargodskiSeiberg}, or models similar to the Kim--Nilles mechanism~\cite{KimNilles} in which $\mu$ arises from a term of the form $\left<S\right>^n/M^{n-1}$~\cite{higherdimmu,YanagidaMu}. In the latter case, one must be slightly careful, e.g. arranging for ${\bar S}^2 S^2/M_P$ terms in the superpotential to be allowed but not simple ${\bar S}S$ masses; discrete gauge symmetries may be arranged to do this. One could also consider stealth sectors with dynamically generated masses (e.g. from confinement) instead of $\mu$ terms in the superpotential. For low-scale breaking, then, one can formulate robust stealth models with the same tools already in use in GMSB models.

However, one of our main concerns in this paper is to show that the stealth mechanism can apply not just for low-scale SUSY breaking, but also for high-scale SUSY breaking. This case requires more care, and we will see that the models are necessarily more elaborate. The basic model-building problem here is that the SUSY-breaking splittings in the stealth sector should be smaller than around 10 GeV\@. Thus, if $m_{3/2} \gsim 10$ GeV, we will need to assume that generic $M_{Pl}$-suppressed operators involving stealth fields are absent; that is, we will have to assume some form of sequestering~\cite{Randall:1998uk}. The simplest realization of this is to sequester both the MSSM fields and the stealth sector together; perhaps they are localized in an extra dimension, far from the source of SUSY breaking. Whether such sequestering is viable in string constructions remains a topic of ongoing research~\cite{sequestering}, but the problem is no worse for stealth models than for more standard phenomenological models, so we will simply assume in this paper that it can be achieved.

Even in sequestered models, there is a remaining technical challenge that will arise in any scenario with gravitinos above the 10 GeV scale. It is roughly analogous to the version of the $\mu/B\mu$ problem arising in anomaly mediation models. Namely, suppose that there is a term in the superpotential with a dimensionful coefficient, like $mX_1 X_2$. Then we should consider $m$ to come with a power of the conformal compensator $\phi = 1 + \theta^2 m_{3/2}$, such that
\beq
{\cal L} \supset \int d^2\theta~m\left(1 + \theta^2 m_{3/2}\right) X_1 X_2 \supset m_{3/2} m X_1 X_2.
\eeq
This is a soft SUSY-breaking $B$-term given by the mass times $m_{3/2}$. Without using the conformal compensator formalism, one can also see such terms directly in supergravity. There will be terms in the superpotential with expectation value $W_0 = m_{3/2} M_P^2$ in order to cancel the cosmological constant, and cross-terms in the $-\frac{3}{M_P^2} \left|W\right|^2$ and $\partial_i W \partial_i K \frac{W}{M_P^2}$ parts of the full supergravity potential will contain the $B$-terms in question. The $B$-terms are very dangerous for stealth SUSY\@. For example, if we have a mass $mS^2$, where $S$ is a stealth field intended to have a supersymmetric mass of 100 GeV and a fermion/scalar splitting of 10 GeV, we have:
\beq
\delta m = m - \sqrt{m^2 - B} \approx \frac{B}{2 m}
\eeq
Then if we have $B = m_{3/2} m$, we require $m_{3/2} \lsim 2 \delta m \lsim 20$ GeV\@. This tells us that in any scenario in which the gravitino is heavy compared to the stealth SUSY splitting, we must ensure that all of the supersymmetric masses for stealth sector fields do not arise from dimensionful parameters in the superpotential. They could, for example, arise from dynamically determined VEVs, FI terms, or confinement.

\subsection{Portals}

Any stealth model must have some way for the lightest $R$-parity odd MSSM particle to decay to the stealth sector, and in return for $R$-even stealth states to decay back to SM particles. A variety of ``portals" exist that can allow these decays. Some of the options are:

\begin{itemize}
\item {\bf Neutral portals}: In this case, a stealth sector operator couples to an operator not charged under any continuous symmetry of the MSSM\@. For instance, given a gauge singlet chiral superfield $S$ in the stealth sector, one can have in the superpotential $\lambda S {\cal O}_{\rm neut}$, where ${\cal O}_{\rm neut}$ is a gauge singlet MSSM operator. Example portals include ${\cal O}_{\rm neut} = H_u H_d$ (the ``Higgs portal"), or ${\cal O}_{\rm neut} = Y{\bar Y}$ where $Y, \bar Y$ are in the ${\bf 5}, {\bf \bar 5}$ representations of the SU(5) GUT group. These portals have been discussed in Ref.~\cite{Stealth}. Other portals include kinetic mixing of U(1) gauge groups, or bifundamental matter charged under both an MSSM gauge group and a stealth gauge group. Many more couplings through higher-dimension operators are available.

Neutral portals might be further subdivided; for instance, $S$ could be a singlet under all symmetries, in which case it will generically have tadpoles which can be problematic in certain models. Alternatively, it might be charged under some nonanomalous discrete symmetries, which can help protect against dangerous divergences as well as explain the presence or absence of various terms in the superpotential.

\item {\bf Charged portals}: The stealth states could be charged under symmetries of the MSSM\@. We don't want them to be charged under gauge symmetries, as gauge interactions would then typically mediate large enough SUSY breaking to produce missing $E_T$ signals. Thus, the reasonable candidates are baryon and lepton number. Portals are associated with $R$-parity-violating terms in the MSSM, with the lowest-dimension examples being $S L H_u$, $S u d d$ (the ``baryon portal"), $S Q L d$, or $S L L E$. Such models suggest the possibility of unifying stealth collider signals with recent attempts to use such portals to explain the abundance of dark matter, baryons, or both~\cite{Kaplan:2009ag,Shelton:2010ta,Davoudiasl:2010am}. The lepton portals generically give rise to final states containing neutrinos, potentially reintroducing missing $E_T$, so we will limit our discussion in this paper to the baryon portal.
\end{itemize}

Notice that in order for the stealth suppression of missing energy to work, we need the proper signs of SUSY-breaking splittings in the stealth sector, which depend on which portal we wish to use. For example, the portal $SH_u H_d$ will always lead to the LOSP decaying to one or more SM particles plus the fermion $\tilde S$, so we need $m_{\tilde S} > m_S$. (More generally, the scalar and pseudoscalar in $S$ may be split, and we only need one of them to be lighter than $\tilde S$.) On the other hand, the operator $Sudd$ lead to the LOSP decaying to SM particles plus the scalar $S$, and in this case we will require that the fermion $\tilde S$ is lighter.

Aside from a choice of portal, we require a mechanism explaining why a decay of a stealth particle to its nearly-degenerate superpartner and a light $R$-odd particle will always (or almost always) occur in any decay chain. Aside from the gravitino, a Goldstone fermion (superpartner of a Goldstone boson) can be a generic candidate for the light $R$-odd field. More general models will often require a symmetry explanation of why certain decays are forbidden; in the case of the baryon portal, baryon number can be useful for this purpose, whereas for neutral portals one might consider discrete gauge symmetries.

Now that we've given a coarse-grained overview of the requirements and challenges involved in building general stealth SUSY models, let us take a closer look at some particular beasts from the zoo of possible models.

\section{Stealth Models with Singlets}
\label{sec:singlet}
\setcounter{equation}{0}
\setcounter{footnote}{0}
In the section, we will discuss a class of models where the stealth sector consists of singlets, either fundamental (in Sec.~\ref{subsec:singlet}) or composite fields (in Sec.~\ref{subsec:vector}). For the most part, we will assume that they communicate with the MSSM through SM vector-like messenger pairs, e.g., transforming as $5 +\bar{5}$ under the SM $SU(5)$.

\subsection{The Vectorlike Portal}
\label{subsec:singlet}
In this section we first review the model in~\cite{Stealth} where the singlet $S$ communicates with the MSSM through messengers $Y, \bary$, which transform as $5 +\bar{5}$ under the SM $SU(5)$.
\begin{equation}
W  \supset  \frac{m}{2} S^2+\lambda S Y \bary + M_Y Y \bary.
\label{eq:SYYsuper}
\end{equation}
Notice that $S$ must be a pure singlet; any symmetry under which it is charged is broken when $\lambda$ and $M_Y$ are both nonzero. Integrating out $Y$ and $\bar{Y}$ with just the supersymmetric mass $M_Y$ at one loop yields the operator
 \beq
c \int d^2 \theta S W_\alpha W^\alpha + h.c.,
\label{eq: dim5op}
\eeq
where $W_\alpha$ is the gauge field strength. The matching coefficient $c$ is 
\beq
c=\sum_i  \lambda \frac{ \sqrt{2}\alpha b_i}{16 \pi M_Y} ,
\eeq 
where the sum runs over the contributions to the gauge coupling beta function coefficients from all heavy fields running in the loop. In the paper we adopt the convention that the kinetic gauge field term is normalized to $-\frac{1}{4} {\rm{Tr}}(F^{\mu\nu}F_{\mu\nu})$.  Writing Eq.~\ref{eq: dim5op} in components, one get operators such as $\lambda^a  \sigma_{\mu\nu}G^{a\mu\nu}\tilde{s}$ and $sG^a_{\mu\nu}G^{a\mu\nu}$. These interactions would induce decays of the gluino to singlino plus gluon and of the scalar $s$ to gluons. Similar operators between $S$ and other SM  vector multiplets exist, which allow decays of neutralinos (charginos) to singlino plus $\gamma/Z$ ($W$) and of $s$ to two $\gamma$'s. These decays are illustrated in Figure~\ref{fig:decaySYY}.

\FIGURE[h]{
\includegraphics[scale=0.8]{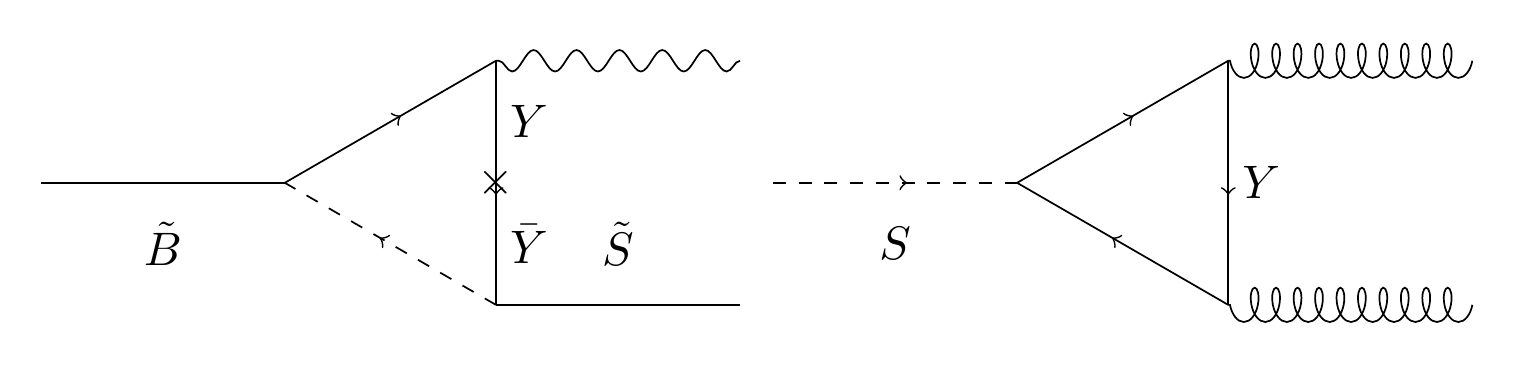}
\caption{The decay of a bino LOSP to a singlino and photon through the vectorlike portal, and the decay of the singlet scalar to gluons.}
\label{fig:decaySYY}
}
 
The $Y$ and $\bary$ contribution to the strong coupling beta function coefficient is 1. Thus the decay widths of the CP even/odd scalar in $S$ to gluons are 
\beq
\Gamma_{s_r(a)}=\frac{\lambda^2\alpha_s^2m_{s}^3}{64\pi^3 m_Y^2},
\eeq
where $\alpha_s$ is the strong coupling. Similarly one could obtain the partial widths to two photons. Such decays are prompt at colliders if $M_Y$ is near the TeV scale (the lifetimes become comparable to those of $B$ mesons or $\tau$ leptons, i.e. of order 100~$\mu{\rm m}$, when the $Y$ mass is about 10 TeV).

\subsection{Tadpole Problems}
\label{subsec:tadpole}
For the superpotential we have presented in~\ref{subsec:singlet}, any possible global symmetry under which $S$ is charged is broken by the coupling $\lambda$. Thus $S$ will obtain a tadpole at one-loop order, which is logarithmically divergent. Given the superpotential~\ref{eq:SYYsuper}, at one-loop order, there are three tadpole diagrams: two involve a scalar ($Y$ or $\bar{Y}$) running in the loop while the third one has the $\psi_Y, \psi_{\bar{Y}}$ fermions in the loop with a supersymmetric mass insertion. The net result, $V_{tadpole}=T S$, is logarithmically divergent by dimensional analysis. An explicit calculation confirms that
\beq
T = - \frac{\lambda m_Y}{(4\pi)^2} \left(6m_{\tilde{D}}^2\log\frac{\Lambda^2}{m_Y^2+m_{\tilde{D}}^2}+4m_{\tilde{L}}^2\log\frac{\Lambda^2}{m_Y^2+m_{\tilde{L}}^2}\right),
\eeq
where $\Lambda$ is the messenger scale where SUSY breaking effect is mediated to the low-lying states. The tadpole will induce a VEV for S\@. Consequently, the masses of both fermion and scalar components in $Y (\bary)$ are shifted:
\beqs
m_{\psi_Y}&=&\lambda \langle S \rangle + m_Y \nonumber \\
m_{Y}^2&=& (\lambda \langle S \rangle + m_Y)^2\pm\lambda m \langle S \rangle+ m_{soft}^2 .
\eeqs
Given $\lambda=0.2, m_{\tilde{D}}=300$ GeV, $m_{\tilde{L}} =200$ GeV, $\Lambda= 100$ TeV, $m_Y$ = 1 TeV, $m$ =100 GeV, we have $\langle S \rangle = 813$ GeV and $\delta m_{\psi_Y} = 162$ GeV\@. One interesting feature is that, because the supertrace in the $Y$ sector is unchanged, the $\left<S\right>$ dependence cancels out of the soft mass for the $S$ scalar:
\beq
m_s^2 \sim -\frac{|\lambda|^2}{(4\pi)^2} \left(6 \tilde{m}_D^2 + 4 \tilde{m}_L^2\right) \log\frac{M_{\rm mess}^2}{m_Y^2}.
\eeq
On the other hand, we have omitted terms that are allowed in the superpotential, such as $\kappa S^3$. In the presence of such terms, a VEV for $S$ could lead to splittings that alter stealth phenomenology. Hence, we must assume either that $\kappa$ is small or the portal coupling $\lambda$ is small, where ``small" in practice means $\simlt 10^{-2}$.

Thus, in gauge mediation scenarios, the $S$ tadpole is not a severe problem for model-building. However, the tadpole can be more troublesome in certain scenarios. Suppose that we wish to consider, instead of a decay ${\tilde S} \to S {\tilde G}$, a decay where another fermion replaces the gravitino. One might consider a model
\beq
W = \frac{m}{2} S^2 + \lambda S Y {\bar Y} + M_Y Y {\bar Y} + \frac{y}{2} S^2 N.
\eeq
In such a model, in the presence of an $S$ tadpole $-TS$, the potential is:
\beq
V \supset \left|y S N + m S + \lambda Y {\bar Y}\right|^2 + \left|\frac{y}{2} S^2\right|^2 - T S.
\eeq
The tadpole leads $S$ to get a VEV, and then $N$ feels a nontrivial potential:
\beq
\frac{\partial V}{\partial N} = y S \left(y S N + m S + \lambda Y {\bar Y}\right)^\dagger~~\Rightarrow~\left<N\right> = -\frac{m}{y}.
\eeq
In this case the VEV for $N$ cancels the leading supersymmetric mass term for $S$. Other models can alter the details, but don't change the basic fact: when $S$ is a pure singlet, it inevitably gets a VEV, in the presence of which an $S^2 N$ superpotential term for $S$ decays will tend to generate a VEV for $N$ which spoils the stealth mass spectrum of $S$. Hence, the $S^2 N$-type models are viable only if $S$ and $N$ can be charged under a symmetry that forbids these tadpole problems; we will discuss such a theory in Section~\ref{sec:Sudd}. 

In fact, the situation is even worse in theories of high-scale SUSY breaking. Naively, one might expect that tadpoles are generated only after SUSY-breaking and are at most of order $m_{\rm soft}^3$. However, as shown in~\cite{Bagger:1995ay}, in general tadpoles are quadratically divergent, and can be of order $m_{3/2}^2 \Lambda^2/M_P$ in a theory with cutoff $\Lambda$ and higher-dimension K\"ahler potential operators suppressed by $M_P$. Taking $\Lambda$ to be of order the Planck scale, we expect that singlets will obtain tadpoles of order $\frac{1}{16\pi^2} m_{3/2}^2 M_P$, possibly with an extra loop factor suppression in theories with extra structure like no-scale breaking. For low-scale SUSY breaking (gravitinos not much heavier than 1 keV) this is not a problem for stealth structure, but for high-scale SUSY breaking it is devastating.

These considerations tell us that, if we want to build stealth singlet models with the same phenomenology as that of Section~\ref{subsec:singlet} that are compatible with high-scale SUSY breaking, we should charge $S$ under a symmetry that can forbid it from obtaining a tadpole.

\subsection{Viable High-Scale Model}
\label{subsec:viable}

We will now consider an explicit model of high-scale SUSY breaking, in outline. (The full details may be found in Appendix~\ref{app:viable}.) As we have just argued, we would like to charge the field $S$ under a symmetry to avoid tadpole problems. We will take this to be a discrete gauge symmetry. We further wish to avoid large $B$-terms, so we would like to start with renormalizable superpotential terms and generate masses dynamically.

The idea, then, is to begin with a superpotential of the form $y SQ{\bar Q} + \lambda SY{\bar Y} + \kappa S^3$, where the $Q$ fields are charged under a new gauge group and the $Y$ fields are ${\bf 5}$ and ${\bf \bar 5}$ fields as before. The gauge group under which the $Q$ fields are charged should have a number of colors and flavors that lead to the dynamical generation of an Affleck-Dine-Seiberg superpotential~\cite{Affleck:1983rr}, $W_{\rm eff} = \Lambda^k/\det({\bar Q}Q)$. Then, as in~\cite{YanagidaMu,Dine:2009swa}, the singlet fields $S$ and the meson fields $Q_i {\bar Q}_i$ will dynamically get VEVs related to the scale $\Lambda$.

In fact, in this model it is possible to arrange for a mesino state, a composite of the fields $Q, {\bar Q}$, to be the lightest $R$-odd fermion, and for a singlino from the $S$ field to decay as ${\tilde S} \to S {\tilde M}$. This decay is prompt on collider timescales. SUSY breaking can be arranged to arise from anomaly mediation. The explanation for stealthiness, then, is just that the AMSB soft terms are proportional to small couplings among the fields we have introduced.

The full model-building requires a bit more care in defining an anomaly-free discrete symmetry, and a particular hierarchy of couplings, $y \ll \kappa \ll 1$. The details have been sequestered in Appendix~\ref{app:viable}.

\subsection{Goldstone Fermion Models}
\label{subsec:goldstone}

One natural possibility is that the fermion in the final state of the stealth decay is light because it is the superpartner of a (pseudo-)Goldstone boson, i.e., it is a Goldstone fermion. Perhaps the most familiar examples of Goldstone fermions are axinos. The effective theory of Goldstone fermions has recently been discussed in~\cite{Higaki:2011bz,Cheung:2011mg,Bellazzini:2011et, Bae:2011jb}. We consider a (dimensionless) chiral superfield $A$ with a transformation $A \to A + i \theta$. Then in the K\"ahler potential, we can consider terms like:
\beq
K \supset \frac{1}{2} f^2 \left(A + A^\dagger\right)^2 + S^\dagger S + c \left(A + A^\dagger\right) S^\dagger S + \ldots,
\label{eq:K axino}
\eeq
where $f$ is the decay constant of the Goldstone. If the broken symmetry is in fact anomalous, so that the Goldstone is an axion, then one might have additional couplings that break the shift symmetry explicitly:
\beq
{\cal L} \supset b \int d^2 \theta~A W_\alpha W^\alpha.
\eeq 
Note that both types of couplings, $c$ and $b$, are of the form that allow a particle to decay to its superpartner and a Goldstone fermion. Because we are considering high-scale SUSY breaking, we must ask whether it is really natural for the fermionic partner of a Goldstone boson to be protected by the same shift symmetry that protects the Goldstone boson itself. As recently emphasized in~\cite{Cheung:2011mg}, in the presence of generic Planck-suppressed operators in the superpotential one expects the Goldstone fermion mass to be at least of the order of the gravitino mass $m_{3/2}$. For high scale SUSY breaking, this is much larger than the stealth splitting we are interested in. However, this is not a special property of Goldstone fermions, but the generic problem that the stealth sector must be protected from large SUSY breaking effects. Thus, we should only consider models in which the Goldstone fermion (and the symmetry breaking sector it descends from) are sequestered from SUSY breaking, so that the leading dangerous operators in the K\"ahler potential are absent.

In order for decays to be prompt on detector timescales, we cannot have $f$ too large. In particular,~\ref{eq:K axino} corresponds to an interaction
\beq
-\frac{ic}{f} S^\dagger {\bar \psi}^A {\bar \sigma}^\mu \partial_\mu \psi^S + c.c. = -c \frac{m_S}{f} S^\dagger {\bar \psi}^A \psi^{\bar S}+c.c.,
\eeq
where the equality holds on-shell in the presence of a mass term $m_S S{\bar S}$. This leads to a decay width of the singlino to scalar plus axino which is given in the limit of massless Goldstone fermion and small stealth splitting ($m_{\tilde S} = m_S + \delta m$) by:
\beq
\Gamma\left({\tilde S} \to S {\tilde A}\right) = \frac{\left|c\right|^2}{4\pi} m_S \left(\frac{\delta m}{f}\right)^2 = \left|c\right|^2 \frac{m_S}{100~{\rm GeV}} \left(\frac{\delta m}{10~{\rm GeV}}\right)^2 \left(\frac{10^9~{\rm GeV}}{f}\right)^2 \frac{1}{25~{\rm cm}}.
\label{eq:lifetime}
\eeq
A QCD axino with $f \sim 10^9$ GeV and a relatively large coupling $c > 1$ could be the invisible particle $\tilde A$. Or, any symmetry broken at a scale below 10$^8$ GeV could give a viable candidate for this Goldstone fermion. Notice that the phase space suppression in the stealth limit for the Goldstone fermion scenario is more mild than in the case of a gravitino, where additional powers are present because the interaction is suppressed by the small stealth SUSY breaking. One also does not want the decay constant to be {\em much} smaller than $10^5$ GeV in the presence of an $A W_\alpha^2$ coupling, to forbid dangerous decays like ${\tilde g} \to g {\tilde A}$ or ${\tilde B} \to \gamma {\tilde A}$ that are more rapid than decays into the stealth sector.

\subsection{Vector-like Confinement Models}
\label{subsec:vector}
Now we consider a more complicated hidden sector charged under a nonabelian gauge symmetry. In the setup, there are pairs of vector-like superfields $Y$ and $\bar{Y}$ transforming as ${\bf 5}+{\bf \bar{5}}$ under the SM $SU(5)$ and also  as ${\bf 2}+{\bf 2}$ under an additional gauged $SU(2)_h$.  The $SU(2)_h$ is more strongly coupled than the SM gauge groups, and thus we can view the SM as weakly gauging the flavor symmetry of the hidden sector. The matter fields $Y$ and $\bar{Y}$ have a large supersymmetric mass $M$. Below the scale $M$, they could be integrated out and we are left with a pure $SU(2)_h$ super Yang-Mills theory. This model might be thought of as a supersymmetric vectorlike confinement model~\cite{Kilic:2009mi}.

 If we assume the 5-plets have a uniform mass $M$ and plug the 1-loop anomalous dimension into the NSVZ beta function, we find that at the fixed point, $\frac{g^2}{16\pi^2} = \frac{1}{15}$, so the coupling is somewhat strong. After integrating out all the fields, the beta function coefficient for pure SU(2) SYM is $b_0 = 6$, and the confinement scale is $\Lambda = M e^{-8\pi^2/(b_0 g^2)} = M e^{-5/4} \approx 0.3 M$. If we want singlet glueballs at the bottom of the SUSY spectrum, near 100 or 200 GeV, this would imply light colored particles from the 5-plets, which decay only through GUT-suppressed operators and thus lead to long-lived, R-hadron-like phenomenology. However, the CMS collaboration already set a lower limit of 620 GeV at 95\% C.L. on the mass of a (semi-)stable stop~\cite{CMS-PAS-EXO-11-022}. We expect such a limit also applies here. We could either raise the stealth supersymmetric scale a bit higher or split the triplets and doublets in the 5-plets (giving them different supersymmetric masses), which could allow further separation between the triplet states and the light glueballs. For instance, if the doublet mass $M_2$ is 300 GeV and the triplet mass $M_3$ is 1 TeV, the one-loop estimate of the confinement scale is $\Lambda = M_3 e^{-5/4} (M_2/M_3)^{1/3} \approx 192$ GeV\@. (However, the coupling is already $g \approx 5$ at $M_2$, so the perturbative estimate is likely subject to large uncertainties.)

 In the SUSY limit, an effective action analysis~\cite{Farrar:1997fn} tells us that in the low energy spectrum, there are two massive chiral supermultiplets with masses $M_+$ and $M_-$. Both of the masses are of order the $SU(2)_h$ confinement scale $\Lambda$, which we take to be of order a few hundred GeV\@. According to~\cite{Farrar:1997fn}, the heavier one consists of two real scalars, which are $SU(2)_h$ gaugino-gaugino bound states (gluinoballs) in the limit of no mass mixing term in the potential and a Weyl fermion, a gaugino-gluon composite. The lighter one has two real scalars, which in the same limit, correspond to parity even/odd glueballs and another fermionic gaugino-gluon composite.\footnote{The analysis of~\cite{Farrar:1997fn} is not supported by a recent lattice simulation~\cite{Demmouche:2010sf}, which orders the glueball and gluinoball supermultiplets in the opposite way. The low-energy spectrum of pure super Yang-Mills remains as an open question. In this paper, we will stick to the analytic analysis in~\cite{Farrar:1997fn} and~\cite{Farrar:1998rm}.}

SUSY breaking is mediated to the hidden sector through the SM fields. Given an SM gluino mass $M_3$, $Y$ and $\bar{Y}$ receive soft masses at one loop while the $SU(2)_h$ gauginos obtain soft masses at the two-loop order,
\beq
m_{\tilde \lambda} \sim \frac{g^2g_3^2M_3}{(16\pi^2)^2}\log\frac{M_{mess}}{\Lambda},
\eeq
where $g_3$ is the SM $SU(3)_c$ coupling and $M_{mess}$ is the SUSY breaking messenger scale. In the low energy spectrum, this will lead to SUSY-breaking splittings within the gluinoball and glueball supermultiplets~\cite{Farrar:1998rm}. Now, in the lighter multiplet, the scalar glueball will be the lightest state, with a heavier fermion and a pseudoscalar as the heaviest component. The splitting would be of order $m_{\tilde \lambda}$ and is naturally very small. 

Loops of the $Y$ and $\bar{Y}$ fields serve as portals for the SM to decay to the hidden sector and vice versa. For instance, if the gluino is the LOSP, it could decay to the hidden gaugino-gluon composites at one-loop. If the gravitino $\tilde{G}$ is the true LSP, the hidden gaugino will decay at tree-level to $\tilde{G} + g_h$, or in the composite degrees of freedom, the gaugino-hidden gluon composite will decay to the hidden glueball plus gravitino. The hidden glueball could decay back to two SM gluons through a box-diagram mediated by $Y~(\bar{Y})$ fields. 

\section{Stealth Through The Baryon Portal}
\label{sec:Sudd}

\subsection{The Portal: Decays to and from the Stealth Sector}

In this section we will consider the case that a field in the stealth sector carries baryon number, and couples to a baryonic operator in the superpotential:
\beq
W \supset \frac{\lambda_{ijk}}{M} u_i d_j d_k S
\eeq
Because the $udd$ coupling is antisymmetric in down quark flavors, the portal can have the quantum numbers of the $\Lambda$ baryon ($uds$); we will refer to it as the ``baryon portal." Note that $S$ has charge $1$ under $U(1)_B$. As we will see shortly, to have stealth phenomenology the scale $M$ is not extremely high, so we will also be interested in UV completions of this operator. We will consider either the $U$-model:
\beq
W_U \supset a_{jk} U d_j d_k + M U\bar{U} + a_i \bar{U} u_i S + m S {\bar S}
\eeq
or the $D$-model:
\beq
W_D \supset b_{jk} D u_j d_k + M D\bar{D} + b_i \bar{D} d_i S + m S {\bar S}.
\eeq
Here $U$ and $D$ have the gauge quantum numbers of $u$ and $d$, respectively, but baryon number $2/3$, and $\bar U$ and $\bar D$ complete them into vectorlike fields. The $a$'s and $b$'s are flavor-dependent coupling constants. Notice that the superfield $S$ contains an $R$-odd scalar and $R$-even fermion. In every case, operators containing three quark superfields implicitly have color contracted with an $\varepsilon$-tensor. We have also added a field ${\bar S}$, with baryon number $-1$, in order to give $S$ a supersymmetric mass. For high-scale models, we will want to replace $m$ with a dynamical value $\left<X\right>$ where the dynamics that give $X$ a VEV may be similar to the model giving rise to the $S$ VEV in Section~\ref{subsec:viable}.

Our first concern is the condition on $\lambda$ and $M$ necessary for decays to happen within the detector. If the LOSP is a squark, it could decay to the $S$ scalar plus two jets: for example, one might have $\tilde{t}_R \to b s S^*$. Other LOSPs will decay through off-shell squarks to three-jet plus scalar $S$ final states, e.g. $\tilde{B},\tilde{g} \to u_i d_j d_k S^*$. We then assume that the scalar $S$ decays to its fermionic partner and a light soft fermion (which will be discussed in more detail in the following subsections). The fermionic $\psi_S$ then decays back through the portal to three jets through a squark-gluino loop. In order to have viable stealth phenomenology, we require that both the LOSP decay into $S$ and the $\psi_S$ decay back to the SM are prompt enough to not create $\met$ signals at colliders.

\FIGURE[h]{
\includegraphics[scale=0.8]{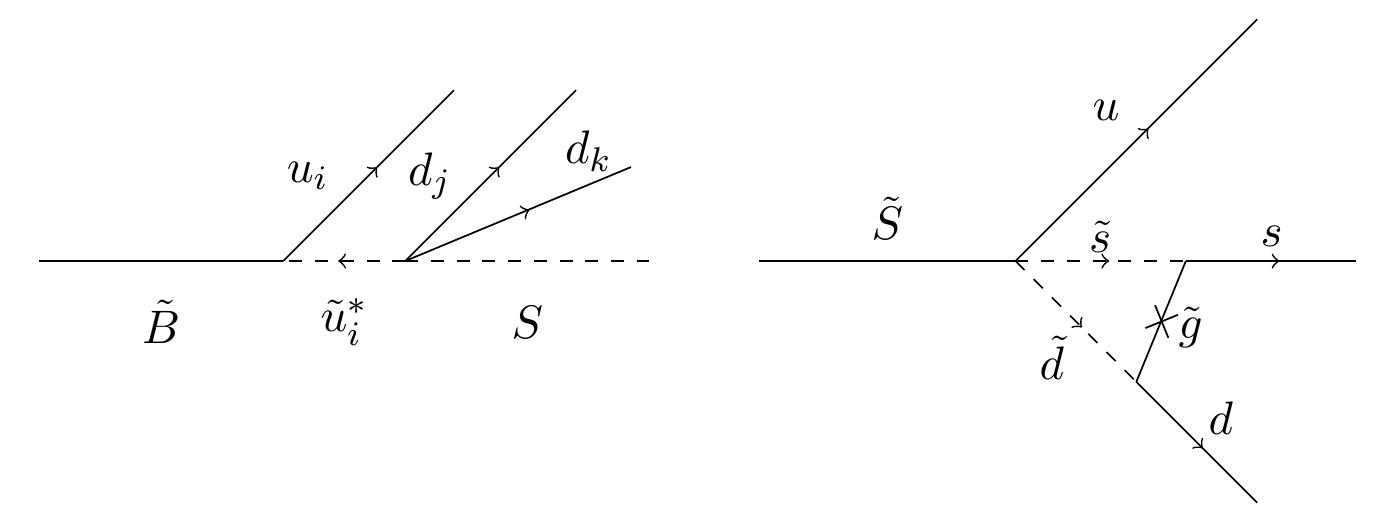}
\caption{Left: one diagram contributing to the decay of a bino LOSP through the baryon portal. Right: a diagram contributing to the loop-level decay of the baryon-charged singlino to 3 quarks.}
\label{fig:binodecay}
}

To be pessimistic about the lifetimes involved, let us assume that the LOSP is a bino. Then it has a 4-body decay $\tilde{B} \to u_i d_j d_k S$, mediated by a dimension-7 operator:
\beq
{\cal L}_{\rm eff} \sim \lambda_{ijk} g' \frac{1}{M m_{\tilde q}^2} {\tilde B}u_i d_j d_k S + h.c.
\eeq
One of the diagrams contributing to this decay is illustrated in Figure~\ref{fig:binodecay}. From this we estimate
\beq
\Gamma \sim \frac{g'^2 \lambda^2}{M^2 m_{\tilde q}^4 (4\pi)^5} m_{\tilde B}^7.
\eeq
However, this is very approximate: aside from order-one factors there are functions of $m_{\tilde q}/m_{\tilde B}$ and $m_S/m_{\tilde B}$ that are important. Thus, we have numerically calculated the lifetime, which is plotted in Figure.~\ref{fig:binolifetime}. (For this calculation we implemented the interactions, with appropriate Lorentz and color structures, in Python in UFO format and used MadGraph5 / MadEvent to compute the widths~\cite{MG5}.) If the bino is relatively light, say below 300 GeV, and squarks are above 1 TeV, the decay is moderate displaced, of order centimeters, for $\lambda = 1$ and $M = 100$ TeV\@. It can be made much more prompt by considering much heavier binos, but in that limit the production of superpartners is likely out of reach of the LHC\@. Thus, we will consider $M/\lambda \simlt 100$ TeV as an upper bound on the scale suppressing the interaction, in the case of a bino LOSP\@.

\FIGURE[h]{
\includegraphics[width=0.57\textwidth]{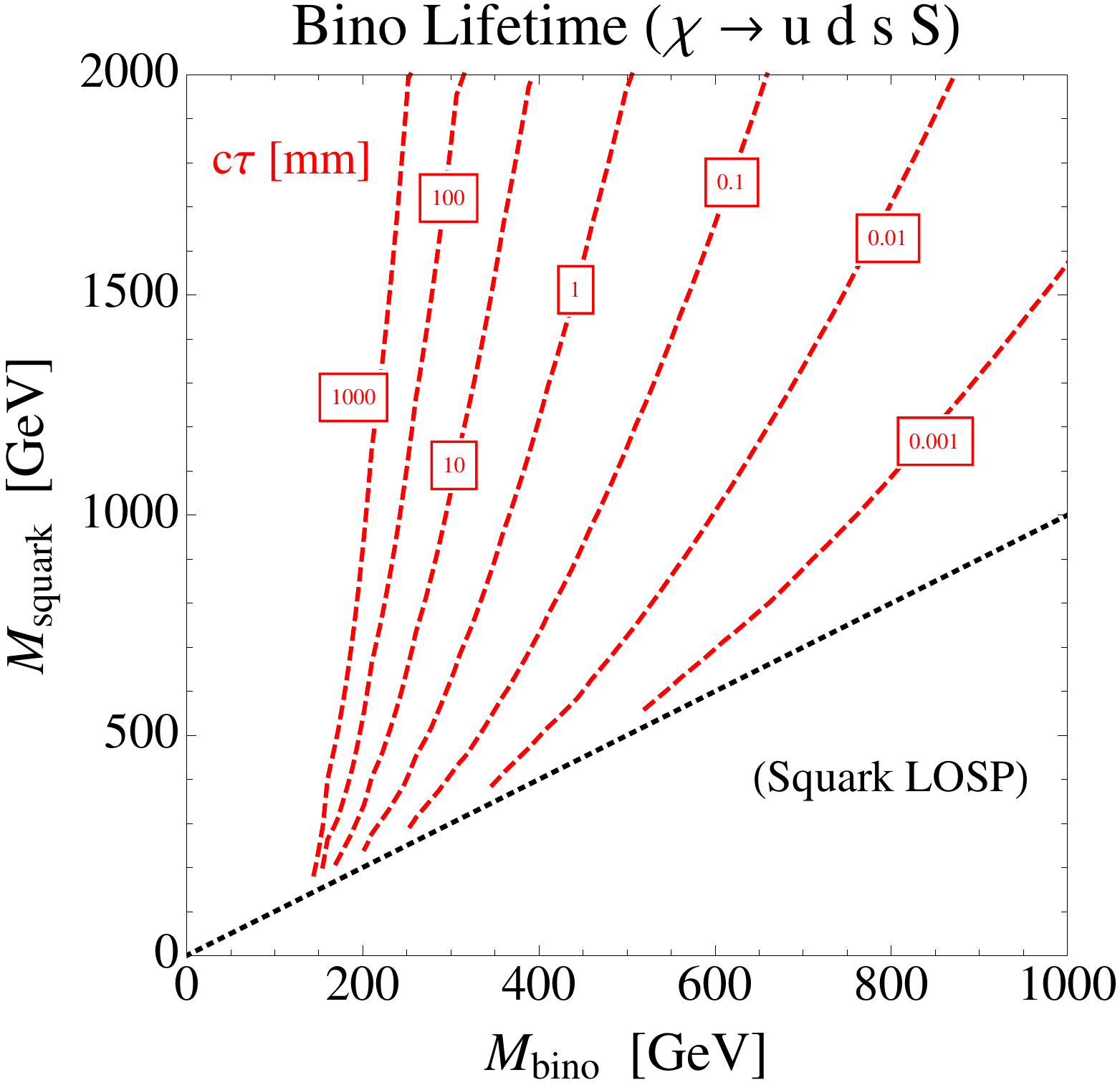} 
\caption{Contours of constant lifetime (in mm) of a bino LOSP decaying through the baryon portal. The assumptions are that $\lambda_{uds} = 1$ with other couplings turned off, the singlet scalar mass is 100 GeV, and $M = 100$ TeV\@. The squark masses are degenerate, $m_{\tilde u} = m_{\tilde d} = m_{\tilde s}$, and given on the vertical axis.}
\label{fig:binolifetime}
}

Less pessimistically, one can consider a squark LOSP decaying directly to the $S$ and two quarks. The squark decay width (for one flavor choice) is:
\beqs
\Gamma(\tilde{u}_i \to d_j d_k S) &=& \frac{\lambda_{ijk}^2 m_{\tilde u_i}^3}{768 \pi^3 M^2} f\left(\frac{m_{S}^2}{m_{\tilde u_i}^2}\right),\\
f(x) & \equiv & 1 + 6 \left(x^2 + x\right) \log x + 9 x - 9 x^2 - x^3.
\eeqs
The phase space factor is important. In particular, if $m_S = m_{\tilde u}/2$, we have $f(1/4) \approx 0.073$, so the decay rate quickly becomes very suppressed relative to the case of massless $S$. Numerically, keeping $m_S = m_{\tilde u}/2$, we have:
\beq
c\tau(\tilde{u}_i \to d_j d_k S) = 6.4 \times 10^{-4}~{\rm \mu m}~\left(\frac{1~{\rm TeV}}{m_{\tilde u}}\right)^3 \left(\frac{M}{100~{\rm TeV}}\right)^2 \frac{1}{\lambda^2}.
\eeq
Thus, a squark LOSP decay is prompt.

The decay {\em out} of the stealth sector is a  $\psi_S$ decay through a squark-gluino loop, illustrated on the right-hand side of Figure~\ref{fig:binodecay}. We find for the decay width:
\beqs
\Gamma &=& \frac{2\alpha_s^2 \lambda^2 m_{\psi_s}^5}{3(4\pi)^5 M^2 m_{\tilde{g}}^2}L^2,
\eeqs 
with $\lambda$ the dominant $\lambda_{ijk}$ and 
\beq
L=\int_0^1dx \int_0^{1-x}dy \frac{1}{x+(1-x)m_{\tilde{q}}^2/m_{\tilde{g}}^2}
\eeq
assuming squarks are degenerate in mass. The decay length is thus, assuming squarks and gluinos have equal masses, 
\beq
c\tau \sim 0.01~{\rm  cm}~\left(\frac{100~{\rm GeV}}{m_{\psi_s}}\right)^5 \left(\frac{M}{10~{\rm TeV}}\right)^2\left(\frac{m_{\tilde{g}}}{1 \, {\rm TeV}}\right)^2 \frac{1}{\lambda^2}.
\label{eq:suddsinglinolifetime}
\eeq
This tells us that, independent of the LOSP, we should not raise $M/\lambda$ far above 100 TeV, and even this is slightly delicate, needing a singlino heavier than 100 GeV or gluinos well below 1 TeV\@.

\FIGURE[h]{
\includegraphics[width=0.56\textwidth]{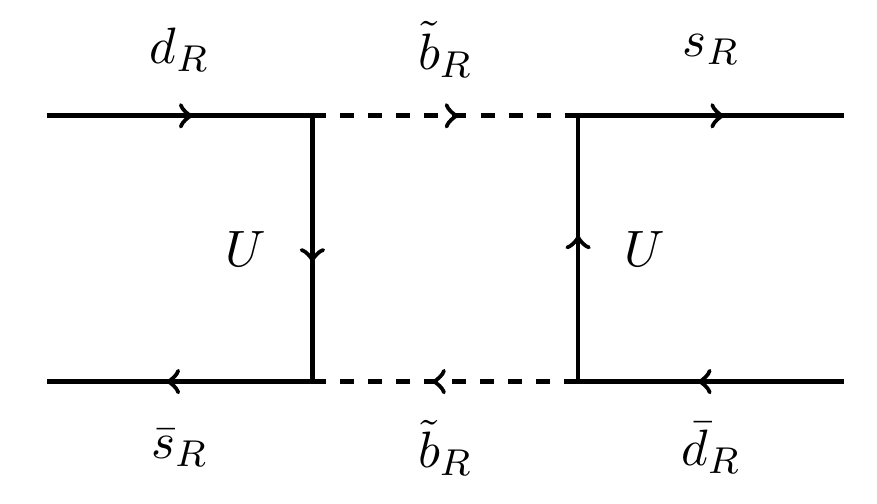} 
\caption{The diagram that contributes to $K^0-{\bar K}^0$ mixing in the $U$-model. A similar diagram exists in the $D$-model, along with other diagrams involving $W$s or winos.}
\label{fig:KKbarmixing}
}

Because the scale suppressing the new physics is 100 TeV or lower, and the $\lambda_{ijk}$ couplings necessarily couple different generations of down (s)quarks, we should ask whether there are strong constraints from flavor physics. The situation is very similar to that of R-parity violation with the $udd$ operator, except that we have not actually violated baryon number, so processes like nucleon-antinucleon oscillations are not allowed. We expect that $K-{\bar K}$ mixing is the most important constraint. In the $U$-model, the contribution is shown in Figure.~\ref{fig:KKbarmixing}, and is similar to RPV contributions discussed in Refs.~\cite{Barbieri:1985ty,Slavich:2000xm}. (The $D$-model has additional contributions, but they don't change the qualitative conclusion.) Comparing to the bound on the corresponding operator $({\bar d}_R \gamma^\mu s_R)({\bar d}_R \gamma_\mu s_R)$ as quoted in Ref.~\cite{Isidori:2010kg}, we find a constraint:
\beq
\frac{M}{a_{db}a_{sb}^*} \simgt 160~{\rm TeV}.
\eeq
This is compatible with our requirements on the lifetime given the above estimates, and suggests that $M$ should be near $100$ TeV if all couplings are order one. (Note that the $\lambda$ couplings are, e.g., $\lambda_{udb} \propto a_{db} a_u$, so the bound doesn't directly constrain them, although we should take $a_u$ small relative to $4\pi$.) The bound can be avoided entirely if, for example, $a_{ds} \gg a_{db,sb}$.

One assumption that might be questioned is the natural size of the couplings $\lambda_{ijk}$. This depends on unknown UV physics. If one assumes MFV, for example, then we should have that $\lambda_{ijk} \propto \epsilon_{abc} (Y_u)_{ai} (Y_d)_{bj} (Y_d)_{ck}$ up to some unknown constant determined by microscopic physics. Assuming the constant is of order 1, even the largest $\lambda$ will be small, of order $10^{-3}$ or less~\cite{Nikolidakis:2007fc}. However, since our $\lambda$'s arise at a scale near the TeV scale from products of couplings with different flavor structure, it is not clear that MFV is a reasonable hypothesis. Besides, even in the MFV framework, the constant could be large, e.g. $\sim 10^3$, without violating any principle. Detailed considerations of flavor physics and the origin of these couplings is beyond the scope of this paper.

\subsection{Baryon Portals with Invisible Gravitino}
\label{subsec:suddgmsb}

Now that we understand how decays into and out of the stealth sector work, we come to the problem of the stealth decay itself. The simplest case is to follow the pattern of the models discussed in Ref.~\cite{Stealth} and consider the decay $S \to \psi_S \psi_{3/2}$, which occurs promptly for low-scale SUSY breaking. For this decay to happen, we require a positive soft mass ${\tilde m}_S^2$. In the context of low-scale SUSY breaking, one usually considers models of gauge mediation (though other possibilities exist). $S$ couples to fields with gauge interactions through either ${\bar U} u_i S$ or ${\bar D} d_i S$ Yukawa couplings, so we can ask whether either of these will automatically produce the correct sign for the $S$ soft mass. The usual RG formulas for soft masses generated from Yukawa couplings tells us that $S$ will obtain a positive soft mass if ${\tilde m}_{\bar U}^2 + {\tilde m}_u^2 < 0$ or, in the other model, ${\tilde m}_{\bar D}^2 + {\tilde m}_d^2 < 0$. Within the framework of general gauge mediation (GGM)~\cite{GGM}, we can obtain different soft masses for $\bar U$ and $u$ only in the presence of an effective FI term for hypercharge. In GGM with $U,\bar U$ or $D,\bar D$ and an effective FI term $\xi_Y$, there are two sum rules (compared to only one in GGM with $\xi_Y$ and without the new fields). One can show that in the $U$-model, achieving ${\tilde m}_{\bar U}^2 + {\tilde m}_u^2 < 0$ implies:
\beq
{\tilde m}_Q^2 - 2 {\tilde m}_d^2 - {\tilde m}_L^2 > 0,
\eeq
which {\em cannot} be satisfied simultaneously with the usual sum rule
\beq
6 {\tilde m}_Q^2 - 9 {\tilde m}_d^2 + 3 {\tilde m}_u^2 - 6 {\tilde m}_L^2 + {\tilde m}_e^2 = 0,
\eeq
without making at least one of $d$, $u$, or $e$ tachyonic. On the other hand, in the $D$-model, the condition for ${\tilde m}_{\bar D}^2 + {\tilde m}_d^2 < 0$ implies:
\beq
{\tilde m}_L^2 - {\tilde m}_Q^2 - {\tilde m}_u^2 > 0,
\eeq
a condition which can be satisfied. Thus, although we have not specified a full model, the GGM formalism tells us that it is reasonable to expect models where the $S$ scalar is heavier than the $S$ fermion, and that they will be associated with effective hypercharge FI terms and heavy sleptons. (The sum rule shows that when this is true, one also has ${\tilde m}_e^2 > 9 {\tilde m}_d^2 + 3 {\tilde m}_u^2$.) This may be an odd corner of parameter space, but there is nothing obviously wrong with it. (Indeed, the apparent preference of this GGM analysis for light squarks is nicely consistent with the sort of spectrum that provides reasonably prompt decays in our model.) One could also consider moving beyond the GGM framework, e.g. by making some SM chiral superfields composite as in Refs.~\cite{singlesector} and related work. Such models can have different spectra of soft masses that are not flavor-degenerate and evade the GGM sum rules.

Aside from the sign of the $S$ soft mass squared, we also have to check its {\em size}. But, as we saw in Ref.~\cite{Stealth}, a splitting appropriate for stealth phenomenology (e.g., $m_S - m_{\psi_S} \approx 10$ GeV) is easily achieved from RG running (here, between the messenger scale and $M$) if the Yukawa coupling ($b_i$, in this case) is $\sim 0.1$. This is perfectly consistent with the requirements for prompt decays and reasonable flavor physics, with $M \approx 10$ TeV and $b_{jk} \sim 1$.

\subsection{Baryon Portals with An Invisible Baryon-Charged Fermion}
\label{subsec:s2n}

One interesting model is for $S$, instead of decaying to its fermionic partner plus another weakly-interacting fermion, to decay to a new fermionic field in the stealth sector that also carries baryon number. This leads us to the superpotential (working in the $U$-model for concreteness)
\beqs
W & = & a_{jk} U d_j d_k + M U\bar{U} + a_i \bar{U} u_i S +\alpha X S{\bar S}+\lambda S^2N \nonumber \\
& \approx & \frac{uddS}{M} + m_S S{\bar S}+\lambda S^2N  \label{eq:fullW}
\eeqs
where the U(1)$_B$ charge of $S$ is $+1$, of $\bar{S}$ is $-1$, and of $N$ is $-2$. Notice that the mass term for $N$ is forbidden by baryon number symmetry and thus $N$ is naturally light. We have replaced a mass term $m_S S {\bar S}$ with a Yukawa coupling $\alpha$ to a field $X$ that we assume obtains a VEV\@. This choice is made to avoid the $B$-term problem of large scalar soft masses $\sim m_S m_{3/2}$; the dynamics which gives $X$ a VEV may be similar to that discussed in Section~\ref{subsec:viable}, relying on gauge theories to generate a dynamical scale. We will simply assume that $\left<X\right> \neq 0$ and that the $X$ field obtains a mass and does not appear in the decay chains we discuss. Thus, the effective theory at low energies is given by the second line of~\ref{eq:fullW}.

Because $N$ is a light field carrying baryon number, it potentially leads to rare baryon number violating processes in normal matter that could have been detected. One concern is double nucleon decay, from $n + n \to N^\dagger+K^0+K^0$ or $n+n \to N^\dagger+(\gamma~{\rm or}~\pi^0)$. Diagrams contributing to such processes can be found by studying the similar baryon-number-violating process in RPV models~\cite{Barbieri:1985ty,Goity:1994dq} and attaching $S$ and $N$ lines to the graphs. Two such diagrams are shown in Figure~\ref{fig:doublenucleon}.

\FIGURE[h]{
\includegraphics[width=0.9\textwidth]{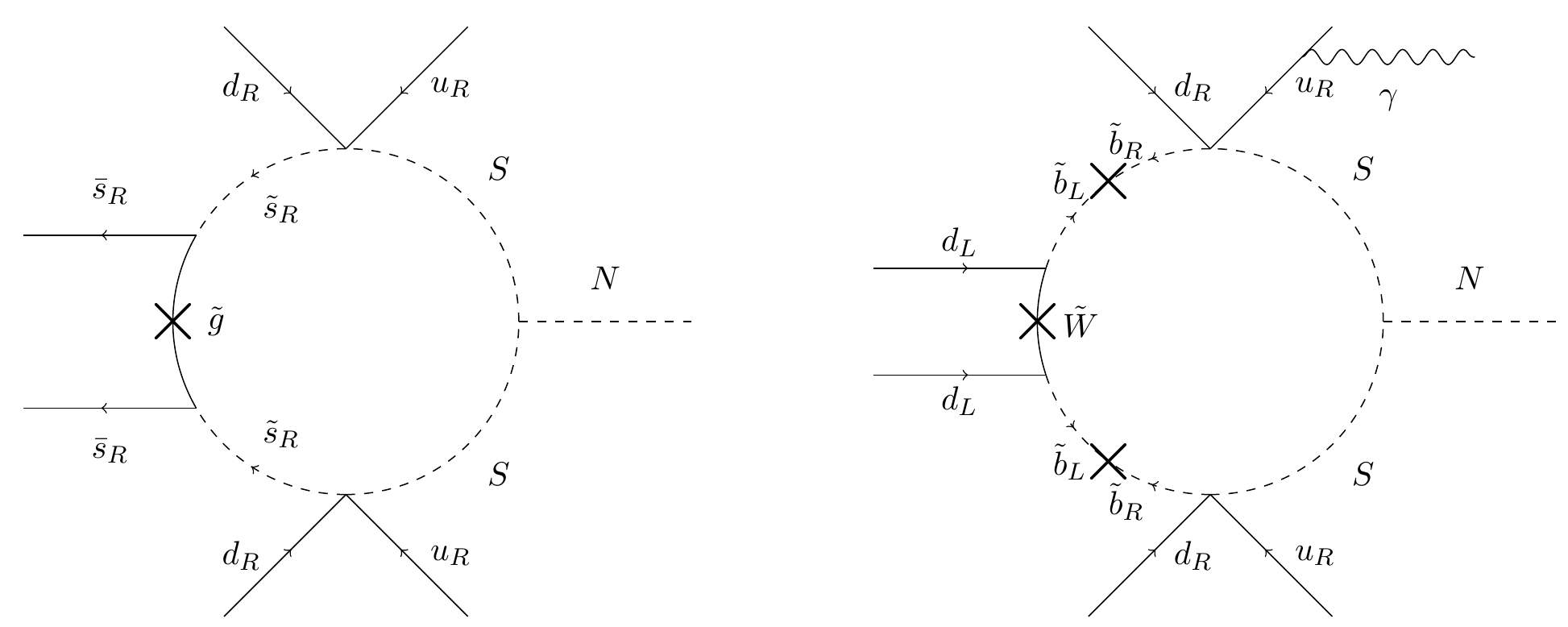}
\caption{Diagrams contributing to double-nucleon decay processes. At left, a contribution to $n + n \to N^\dagger + K^0 + K^0$, a channel that is open if $m_N < 884$ MeV\@. At right, a contribution to $n + n \to N^\dagger + \gamma$, open if $m_N < 1879$ MeV\@. Notice that the latter process requires insertions of flavor violation and left-right squark mass mixings. Both diagrams require an insertion of a Majorana gaugino mass.}
\label{fig:doublenucleon}
}

The $n + n \to N^\dagger + K^0 + K^0$ decay goes through a dimension-10 operator, which we estimate (up to an order one factor) in the limit in which $m_{{\tilde s}_R} \gg M_{\tilde g} \gg m_S$:
\beq
\sim \frac{g_s^2 \lambda_{uds}^2}{16\pi^2 m_{{\tilde s}_R}^4 M_{\tilde g} M^2} a_\lambda \log\frac{M_{\tilde g}^2}{m_S^2} \left(uds\right)\left(uds\right)N.
\eeq
In the case of just an $R$-parity violating $\lambda_{uds}$ coupling, a crude estimate of the bound from double nucleon decay is~\cite{Goity:1994dq} (taking the squarks and gluinos to be at around 1 TeV) $\lambda^{(RPV)}_{uds} \simlt 3 \times 10^{-5} \left(100~{\rm MeV}/\tilde\Lambda\right)^{5/2}$, where ${\tilde \Lambda}$ is a hadronic scale arising from nuclear matrix elements and other order-one slop in the calculation. Relative to that calculation, the decay amplitude for us is suppressed by a loop factor and $\Lambda^2$ (and by the fact that this is a $2\to2$ rather than $2\to3$ process), and proportional to the $A$-term and a logarithmic factor. Taking $\Lambda = 10$ TeV and $a_\lambda = 10$ GeV, we can obtain a similar estimate in the stealth case: $\lambda_{uds} \simlt 5 \left(100~{\rm MeV}/\tilde \Lambda\right)^3$. The hadronic scale ${\tilde \Lambda}$ appearing in this estimate need not be precisely identical to the one in~\cite{Goity:1994dq}, but we expect it to be of similar size. If it is 100 MeV, there is really no bound at all; if it is somewhat bigger, there may be a very weak bound on $\lambda$, but not one that conflicts with having prompt enough decays for stealth phenomenology unless ${\tilde \Lambda} \gg 1$ GeV\@. In that case, we would require either that some other flavor choices of $\lambda_{ijk}$ are dominant or that the scalar $N$ mass is large enough to forbid the decay.

In the model as written, the $N$ field is massless. One could add a baryon-number violating mass $m N^2$, which breaks baryon number to ${\mathbb Z}/12$. This could be taken small, but heavy enough that $N$ does not influence BBN, for instance. Such a choice is technically natural if this mass is the largest source of baryon-number violation. The remaining ${\mathbb Z}/12$ symmetry is enough to forbid proton decay and neutron-antineutron oscillations.

Now, let us discuss SUSY breaking effects. Again, we need a positive scalar soft mass-squared for $S$ for stealth phenomenology, which could be achieved with low-scale SUSY breaking as discussed in Section~\ref{subsec:suddgmsb}. Alternatively, one could try to build a model using anomaly mediation, similar to that of Section~\ref{subsec:viable}. It turns out that the minimal attempt to do so yields a spectrum that is only partially stealthy, as discussed in Appendix~\ref{app:s2n}; slightly more complicated models would be necessary to give viable high-scale SUSY breaking models of the baryon portal with $S^2 N$ term. Of course, one can also consider models with alternative decays like the Goldstone fermion discussed in Section~\ref{subsec:goldstone}.

\section{Stealth Models with a Light $Z^\prime$}
\label{sec:zprimesector}
\setcounter{equation}{0}
\setcounter{footnote}{0}
In this section, we will consider two types of models in which the stealth sector is charged under a new gauged $U(1)^\prime$ symmetry with a gauge coupling $g_d$. In the first model, $U(1)^\prime$ is broken by an $F$-term potential. In the second model, $U(1)^\prime$ is broken by a $D$-term potential involving a non-zero FI term. 

\subsection{$U(1)^\prime$ Broken by a Superpotential}
\subsubsection{Decays to and from the Stealth Sector}
 In addition to the MSSM superpotential, our model has the following superpotential,
\beq
W=y S (\phi_+\phi_- - \Lambda^2),
\eeq
where the chiral superfields $\phi_+, \phi_-$ have opposite charges $q, -q~(q>0)$ under $U(1)^\prime$ while $S$ is not charged. The scale $\Lambda$ will be taken to be around the weak scale and could be generated dynamically through the retrofitting mechanism~\cite{retrofitting}. 

For this sector to communicate to the MSSM, (some) SM fields (and possibly SM-charged spectators to cancel anomalies) are also charged under this $U(1)^\prime$. One would then immediately worry whether such a low scale $Z^\prime$ sector is already ruled out experimentally. First, to avoid stringent constraints on leptonic processes, this $Z^\prime$ has to be leptophobic or have very tiny couplings to the leptons. For a leptophobic $Z^\prime$, the main constraints come from the dijet resonance searches at hadron colliders and electroweak precision tests (EWPT). 
The strongest limits on a $Z^\prime$ decaying into dijets with mass below 500 GeV actually come from a combination of UA2~\cite{Alitti:1993pn} and CDF results~\cite{Aaltonen:2008dn} (LHC dijet searches are less sensitive to low mass resonances). At the Tevatron's energy $\sqrt{s} = 1.96$ TeV, a $Z^\prime$ produced at a rate of a few tens of picobarns is allowed across the whole range below 500 GeV and a rate of hundreds of picobarns is allowed below $m_{Z^\prime} = 350$ GeV\@. For an order one coupling $g_d \sim 1$,  the charge of the first generation quarks under this $U(1)^\prime$ has to be small $\lesssim  0.1$. This is naturally achievable if the first generation quarks only obtain a non-zero charge of order $\sim 0.1$ after mixing with a heavy vector-like set of quarks with order one charge under the $U(1)^\prime$~\cite{effectivezprime}. This effective $Z'$ point of view also relieves us of the need to carefully construct anomaly-free symmetries. Aside from direct constraints, $Z^\prime$ gauge boson mixing with the SM $Z$ boson provides a positive contribution to the oblique $T$ parameter and thus is constrained by the EWPT\@. Parametrizing the off-diagonal component in the $2 \times 2$ mass matrix of $Z$ and $Z^\prime$ as $\gamma m_Z^2$, the correction to the $T$ parameter is then $\alpha T = \gamma^2 \frac{m_Z^2}{m_{Z^\prime}^2-m_Z^2}$ with $\alpha$ the fine structure constant. Fixing $m_h =125$ GeV,  $m_t = 173$ GeV and $U = 0$, the global electroweak fit gives $S = 0.07 \pm 0.09$ and $T  = 0.10 \pm 0.08$~\cite{Goebel:2010ux}, which implies $\gamma \frac{m_Z}{\sqrt{m_{Z^\prime}^2-m_Z^2}} \lesssim 0.036$. If our Higgs is charged under this $U(1)^\prime$, $\gamma = \frac{g_d g Q_{h} v^2}{m_Z^2}$ where $g$ is the SM $SU(2)_W$ coupling and EWPT requires the Higgs charge under $U(1)^\prime$ $Q_h \lesssim 0.03$ for $m_{Z^\prime} = 200$ GeV\@.\footnote{For a two Higgs doublet model with a large $\tan\beta$, the down type Higgs could have a charge of order one as its contribution to the mass mixing of $Z$ and $Z^\prime$ is suppressed by $\tan^2 \beta$~\cite{Fan:2011vw}.} In summary, the experimental data only constrains the charges of the first generation quarks and the Higgs. 

At the scale $\Lambda$, $U(1)^\prime$ is spontaneously broken by the VEVs of the $\phi$ fields $\langle \phi_+ \rangle = \langle \phi_- \rangle \equiv f$. Ignoring SUSY breaking for the moment, $f=\Lambda$. All chiral superfields ($S,~\phi_+,$ and $\phi_-$) obtain masses. The fermion in the linear combination $\phi_1\equiv (\phi_++\phi_-)/\sqrt{2}$ obtains a Dirac mass $m_D=\sqrt{2}yf$ with the fermion in $S$, from the superpotential. The orthogonal combination $\phi_2\equiv(\phi_+-\phi_-)/\sqrt{2}$ is ``eaten" by the super-Higgs mechanism to make the whole gauge multiplet massive with mass $m_g=2 g_dqf$. More specifically, the $U(1)^\prime$ gaugino $\tilde{Z}^\prime$ marries the combination $\tilde{\phi}_2 \equiv (\tilde{\phi}_+-\tilde{\phi}_-)/\sqrt{2}$ to obtain a Dirac mass $m_g$. The imaginary part of $\phi_2$ is eaten by the gauge boson while the real part of $\phi_2$ gets a mass $m_g$ from the $D$-term potential. 

A squark LOSP charged under $U(1)^\prime$ could decay to the quark and the $U(1)^\prime$ gaugino, $\tilde{q} \to q \tilde{Z}^\prime$, while a $U(1)^\prime$-charged Higgsino LOSP could decay to the Higgs and the $U(1)^\prime$ gaugino, $\tilde{H} \to h \tilde{Z}^\prime$. Other LOSPs not charged under the $U(1)^\prime$ could decay through charged MSSM particles, e.g. off-shell squarks, to two jets plus a $\tilde{Z}^\prime$. After SUSY breaking which we discuss in the next section, a small mass splitting is generated between different components of the heavy vector multiplet. Depending on the split mass spectrum, $\tilde{Z}^\prime$ could decay as $\tilde{Z}^\prime \to Z^\prime  + \tilde{X}$ with $\tilde{X}$ the soft invisible fermion or to ${\rm Re}(\phi_2)$ via the mixing with the $\tilde{\phi}_2$, $\tilde{Z}^\prime \to {\rm Re}(\phi_2) + \tilde{X}$. The $Z^\prime$ would subsequently decay back to two jets, $Z^\prime \to q \bar{q}$. For the scalar ${\rm Re}(\phi_2)$, the trilinear scalar coupling from the $D$-term would induce its decay through an off-shell squark-gluino loop to two quarks, ${\rm Re}(\phi_2) \to q\bar{q}$. These possible decay paths are illustrated in figure~\ref{fig:Zpdiag}. All decays in and out of the stealth sector are prompt.

\FIGURE[h!]{
\includegraphics[width=1\textwidth]{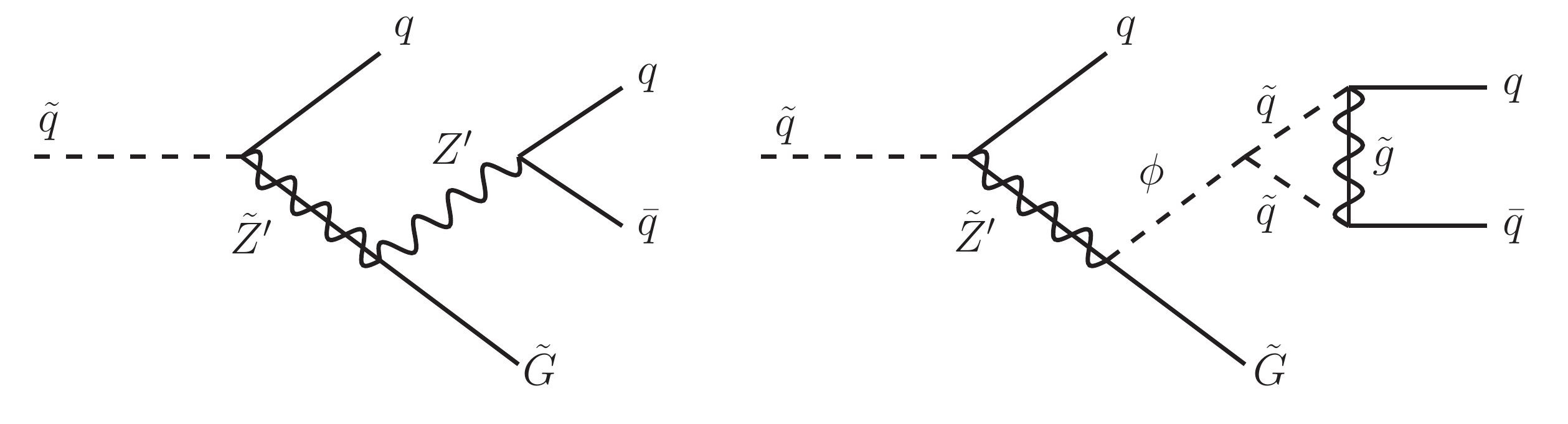} 
\caption{ \label{fig:Zpdiag}
Possible stealth decays through the $Z'$-portal.  The decay on the left passes through the $Z'$, while the decay on the right passes through the scalar within the massive vector supermultiplet.}
}

\subsubsection{SUSY Breaking: Decays inside the Stealth Sector}
Now we consider the effects of SUSY breaking in the stealth sector. We will proceed in three steps: we will first parametrize SUSY breaking by three phenomenological parameters, the $Z^\prime$ gaugino soft mass $m_{\tilde{Z^\prime}}$ and the $\phi_{+(-)}$ soft masses $m_{+(-)}$ and show that they have to satisfy certain conditions to fulfill the stealth mechanism. Then we will check how to satisfy these conditions in different SUSY mediation mechanisms. Finally we discuss the decay inside the stealth sector during which a light invisible particle carries away missing energy.

After SUSY is broken, the whole scalar potential is  $V=|F|^2+|D|^2+m_+^2 |\phi_+|^2+m_-^2 |\phi_-|^2$. For simplicity, we will assume $m_+ = m_- \equiv \tilde{m}_s$, which is true in a broad class of SUSY breaking mediation schemes that are insensitive to the sign of the charge. The minimum is shifted to $f =\sqrt{\Lambda^2-\tilde{m}_s^2/y^2}$. The uneaten scalars' masses are 
\beqs
m_{Im(\phi_1)}&=&\sqrt{2}y\Lambda, \quad m_S= m_{Re(\phi_1)}=\sqrt{2(y^2\Lambda^2-\tilde{m}_s^2)}\approx \sqrt{2}y\Lambda-\frac{\tilde{m}_s^2}{\sqrt{2}y\Lambda},  \nonumber \\
m_{Re(\phi_2)}&=& 2\sqrt{g_d^2q^2\Lambda^2-\left(\frac{g_d^2q^2}{y^2}-\frac{1}{2}\right)\tilde{m}_s^2}\approx 2g_dq\Lambda-\left(\frac{g_dq}{y^2}-\frac{1}{2g_dq}\right)\frac{\tilde{m}_s^2}{\Lambda},
\eeqs
where we expand to the leading order in $\tilde{m}_s^2/\Lambda^2$. The gaugino soft mass $m_{\tilde{Z^\prime}}$ would push one fermion mass eigenvalue down and the other one up. Analytically the fermion masses are 
\beqs
m_{\tilde{\phi}_1}&=&m_{\tilde{S}}=  \sqrt{2(y^2\Lambda^2-\tilde{m}_s^2)}\approx\sqrt{2}y\Lambda-\frac{\tilde{m}_s^2}{\sqrt{2}y\Lambda} , \nonumber \\
m_{\psi_{1(2)}}&\approx&2g_dq\Lambda-\frac{g_dq}{y^2}\frac{\tilde{m}_s^2}{\Lambda}\pm \frac{m_{\tilde{Z^\prime}}}{2},
\eeqs
where $\psi_{1(2)}$ are mixtures of $\tilde{Z}^\prime$ and $\tilde{\phi}_2$. 
The gauge boson mass $Z^\prime$ is 
\beq
m_{Z^\prime}=2g_dqf=2g_dq\sqrt{\Lambda^2-\frac{\tilde{m}_s^2}{y^2}} \approx 2g_d q \Lambda - \frac{g_dq}{y^2}\frac{\tilde{m}_s^2}{\Lambda}.
\eeq
Neglecting the $S, \phi_1$ sector, we see that the mass of the gauge sector, to the leading order in $\tilde{m}_s^2/\Lambda^2$, is shifted by an overall $-\frac{g_dq}{y^2}\frac{\tilde{m}_s^2}{\Lambda}$ after SUSY is broken. Another feature is that the $Z^\prime$ mass is always in between the two fermions $\psi_1$ and $\psi_2$ with $|m_{Z^\prime}-m_{\psi_{1(2)}}| \sim |m_{\tilde{Z^\prime}}|$. The lightest state is a boson, $Re(\phi_2)$, only when the following conditions are satisfied:
\beq
\tilde{m}_s^2 < 0, \quad \frac{|\tilde{m}_s^2|}{g_dq\Lambda} \gg |m_{\tilde{Z^\prime}}|. 
\label{eq: condition}
\eeq
The first condition requires that the scalar get a negative soft mass while the second one tells us that the mass splitting between the scalar and fermions is much larger than that between the gauge boson and fermions.

Now we will look for mediation mechanisms that generate a soft mass spectrum satisfying Eq.~\ref{eq: condition}. One irreducible SUSY breaking contribution to the stealth sector is $U(1)^\prime$ gauge mediation through the visible sector. The gaugino $\tilde{Z}^\prime$ receives soft masses at one loop with the MSSM fields (or the spectators that cancel the $U(1)^\prime$ anomaly) running in the loop. As the gaugino soft mass breaks $R$ symmetry, there must be an insertion of a chirality-flipping scalar soft mass, which breaks $R$ symmetry, in the loop. Assuming contributions only from the MSSM fields, the $\tilde{Z}^\prime$ soft mass is
\beq
m_{\tilde{Z}^\prime}=\frac{3g_d^2q_t^2m_t}{8\pi^2 m_{\tilde{t}}^2}(v a_t \sin\beta-\mu m_t \cos \beta)\log{\frac{m_{\tilde{t}}^2}{m_t^2}}-\frac{g_d^2q_h^2\mu\sin2\beta}{16\pi^2}\log{\frac{m_A^2}{\mu^2}},
\label{eq:gauginosoft}
\eeq
where $v$ is the Higgs VEV, $q_t~(q_h)$ is the charge of the top quark (Higgs) under the $U(1)^\prime$, and $\beta$ is the mixing angle between up- and down-type Higgs. $\mu$ is the dimension one parameter in the MSSM Higgs superpotential and $m_A$ is the mass of the heavy Higgs scalars and pseudoscalar in the decoupling limit. $a_t$ is the top $A$ term and $m_{\tilde{t}}$ is the stop mass where we neglect the mass differences between left and right handed stops. The contributions from other quark supermultiplets are negligible as they always proportional to the fermion masses. 

Now we consider the scalar soft masses from $Z^\prime$ mediation. The fields $\phi_+$ and $\phi_-$ obtain equal soft masses at the two loop order, with MSSM fields as messengers.\footnote{We have assumed here that the one-loop contributions to the scalar soft masses proportional to the trace over the $U(1)^\prime$ charged fields, ${\rm Tr}(q \,m_{soft}^2)$, vanish. This is true in general gauge mediation without a non-zero FI term.} The MSSM messenger mass matrix has a positive non-zero supertrace, which leads to a {\em negative} and logarithmically divergent two-loop contribution to the soft masses of both $\phi_+$ and $\phi_-$. The leading logarithmic contribution from one multiplet labeled by $i$ is~\cite{Poppitz:1996xw}
\beq
\tilde{m}_s^2\approx-\frac{g_d^4 q^2N_i q_i^2}{32 \pi^4}\, {\rm STr}\, m_i^2\,\log{\frac{\Lambda_{UV}^2}{\tilde{m}_i^2}},
\eeq
where $N_i$ counts the degrees of freedom for species $i$, e.g. $N=3$ for a particle transforming as $(3, 1)$ under the SM $SU(3)_c \times SU(2)_W$; $ {\rm STr} \, m_i^2$ is the supertrace of the messenger mass matrix; $\tilde{m}$ is the mass of the scalar messenger; and the UV cutoff $\Lambda$ is the scale where the soft masses of MSSM fields are generated. For $g_d=1, q=1, N_i=3, q_i=0.8, \tilde{m}=400$ GeV and $\Lambda_{UV}=$1000 TeV, we find $\tilde{m}_s^2\approx - (40\, {\rm GeV})^2$. For the supersymmetric mass $m_g =100$ GeV, this leads to a splitting of 8 GeV between scalar and fermions inside the stealth sector, neglecting $m_{\tilde{Z}^\prime}$.

According to Eq.~\ref{eq: condition}, the gaugino soft mass has to be more suppressed compared to the scalar soft mass. In gauge mediation, this is easily achievable as the $A$ terms are generically small and $m_{\tilde{Z}^\prime}$ calculated from Eq.~\ref{eq:gauginosoft} is suppressed. There could be other SUSY breaking contributions to the stealth sector, for instance, gravity mediation in scenarios with a heavy gravitino. To maintain the desirable stealth spectrum, besides sequestering the scalar soft masses $\tilde{m}_s^2$, we require that $\tilde{m}_s^2$ remain negative and the mediation mechanism has to be approximately $R$ symmetric. We will not go further in building complicated high-scale mediation models to realize this but only mention that there is no no-go theorem against these requirements and there are already viable high-scale mediation models with an approximate $R$ symmetry in the literature~\cite{Kribs:2010md}.

\FIGURE[h]{
\includegraphics[scale=0.7]{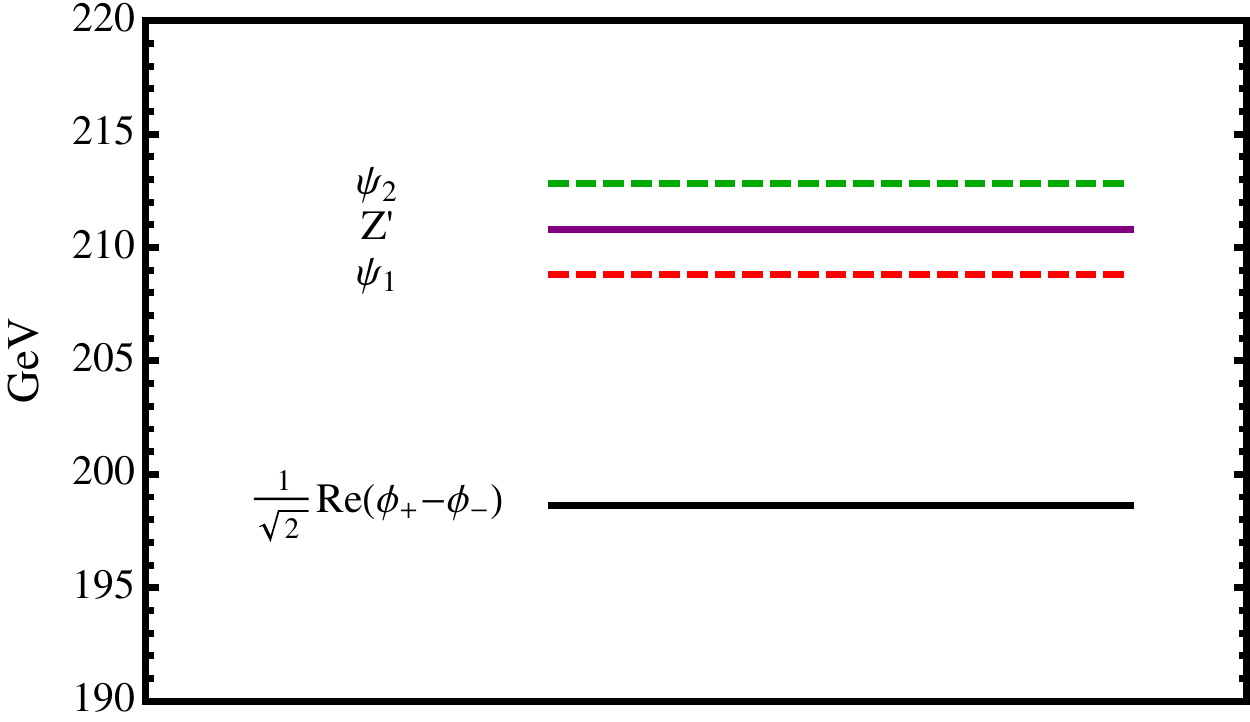} 
\caption{One sample spectrum of the $Z^\prime$ model. Dashed lines correspond to two fermion states, which are mixures of $U(1)^\prime$ gaugino and $(\tilde{\phi}_+-\tilde{\phi}_-)/\sqrt{2}$ fermion. We only show states in the gauge sector, which are relevant for the decays. States such as $S$ and $\phi_++\phi_-$ do not mix with the other states and will not appear in the decay chain. }
\label{fig:stealthspectrum}
}

Finally we put in the invisible particle and estimate the decay lengths of the decay inside the stealth sector. In gauge mediation, the natural candidate for the invisible particle is the gravitino. The decay $\psi_{1(2)} \to Re(\phi_2) + \tilde{G}$ has a decay length ranging from mm to cm depending on the SUSY breaking scale $\sqrt{F}$ as long as the splitting between the fermion and scalar is about 10 GeV\@. The heavier fermion $\psi_2$ could also decay to the gauge boson, $\psi_2 \to Z^\prime (Z^{\prime*}) + \tilde{G}$. As discussed in~\cite{Stealth}, in decays $A \to B + \tilde{G}$, if $\Gamma(B) \sim {\cal O} (1)$ GeV, the width of the three-body decay through an off-shell $B$ will be comparable to that of the two-body decay through an on-shell $B$ particle. As the $Z^\prime$ width could be big, e.g.\ $\Gamma(Z^\prime) \sim 1$ GeV given that the charges of all the light quarks are 0.1, the width of $\psi_2 \to Z^{\prime*} + \tilde{G}$ is comparable to the two-body one and thus this channel is not completely stealthy! On the other hand, as the scalar $Re(\phi_2)$ decays through a loop to two SM quarks and its width is much smaller than 1 GeV, the decay $\psi_{1(2)} \to Re(\phi_2) + \tilde{G}$ is dominantly two-body. Luckily as the mass splitting between the gauge boson and gaugino determined by $m_{\tilde{Z}^\prime}$ always has to be smaller than that between $Re(\phi_2)$ and fermions, e.g. by a factor around 10, the partial width of the channel through gauge boson is suppressed and thus negligible. One natural replacement for the light gravitino is the Goldstone fermion, as discussed in Sec. 3.4, as long as it is light (say, $m_{\tilde{A}} < 10$ GeV) and cosmologically safe. $A$ could couple to the $\phi_+$ and $\phi_-$ fields in the K\"ahler potential as $K\supset(A+A^\dagger)\phi_+^\dagger \phi_+ + (A+A^\dagger)\phi_-^\dagger \phi_-$, and the estimate of the decay length is the same as in Eq.~\ref{eq:lifetime}.

\subsection{Light $Z^\prime$ from Kinetic Mixing}

In Section~\ref{subsec:viable}, we discussed the use of a strongly coupled gauge theory to generate a dynamical scale. An alternative mechanism for dynamically producing a mass term is to generate a VEV using kinetic mixing of U(1) gauge groups, $-\frac{1}{2}\epsilon\int d^2\theta {\cal W}_\alpha {\cal W}'^\alpha$. This is the supersymmetrization of the phenomenon studied in~\cite{Holdom}. As shown in~\cite{Cheung:2009qd} (also see~\cite{Dienes:1996zr}), it can generate a supersymmetric mass scale in a sector with no explicit superpotential mass terms, starting from a U(1) in a different sector with a $D$-term expectation value. This will produce a VEV $v^2 = \epsilon\left<D\right>$. We can first ask whether a hypercharge $D$-term VEV can suffice. The experimental limits on kinetic mixing with a new light $Z'$ are typically at least as strong as $\epsilon \simlt 0.03$ for most $Z'$ masses below or of order 100 GeV~\cite{Hook:2010tw}. Suppose we take, for example, $D_Y = \left(2~{\rm TeV}\right)^2$, $\epsilon = 0.03$, a dark U(1) coupling $g_d = 0.3$, and a dark sector superpotential $W = \lambda S \phi_+ \phi_-$ with $\lambda = 0.2$. Then the $D$-term potential $g_d^2 \left(\left|\phi_+\right|^2 - \left|\phi_-\right|^2 - \xi_d\right)$, with $\xi_d = \epsilon D_Y$, will lead to a VEV $v_d = \left<\phi_+\right>$, producing multiplets of dark photon states at mass $g_d v_d \approx 100$ GeV and of dark Higgs states at mass $\lambda v_d \approx 70$ GeV\@.  As in~\cite{Cheung:2009qd}, we expect a scalar dark Higgs to be lighter than its fermionic partner if kinetic mixing is the dominant mediation of SUSY breaking to the dark states. (One could also consider AMSB, arranging for $g_d$ large enough relative to $\lambda$ to produce a similar spectrum.) The dark Higgs decays through two off-shell dark photons; using the results in~\cite{arXiv:0903.0363}, we can see that for the relatively heavy dark Higgs mass and kinetic mixings that we consider, this decay is prompt, unlike in previous phenomenology. The cascade decays proceed much as in the study of lepton jets~\cite{LeptonJets}, with a bino decaying to dark Higgs and dark Higgsino, ${\tilde B} \to {\tilde h}_d h_d$, followed by ${\tilde h}_d \to h_d {\tilde G}$ given low-scale SUSY breaking or perhaps to a Goldstone fermion in the case of high-scale breaking. Dark Higgses then produce jets of charged quarks and leptons; because of the mass scales we consider, these will no longer be primarily lepton jets. Similar decays through a dark sector already have been known to hide SUSY; here, the challenge is made even greater by combining this with the stealth mechanism.

Alternatively, one could consider models in which the portal to the MSSM is independent of kinetic mixing, but kinetic mixing with a new heavy gauge boson generates a supersymmetric mass scale in the stealth sector. For instance, in models of low-scale SUSY breaking, it may be that new U(1) gauge groups exist at the messenger scale. Then a relatively small kinetic mixing, such as the $10^{-3}$ to $10^{-4}$ mixings that we expect to be typically generated by loops, can lead to weak-scale FI terms for a stealth sector U(1), inducing VEVs that can be used to generate stealth-sector masses. This provides an interesting alternative to the nonabelian gauge dynamics needed to generated masses along the lines of Section~\ref{subsec:viable}, while preserving the portals and collider signatures of such models.

\section{Stealth at the LHC}
\label{sec:collider}
\setcounter{equation}{0}
\setcounter{footnote}{0}

\begin{figure}
\includegraphics[width=\textwidth]{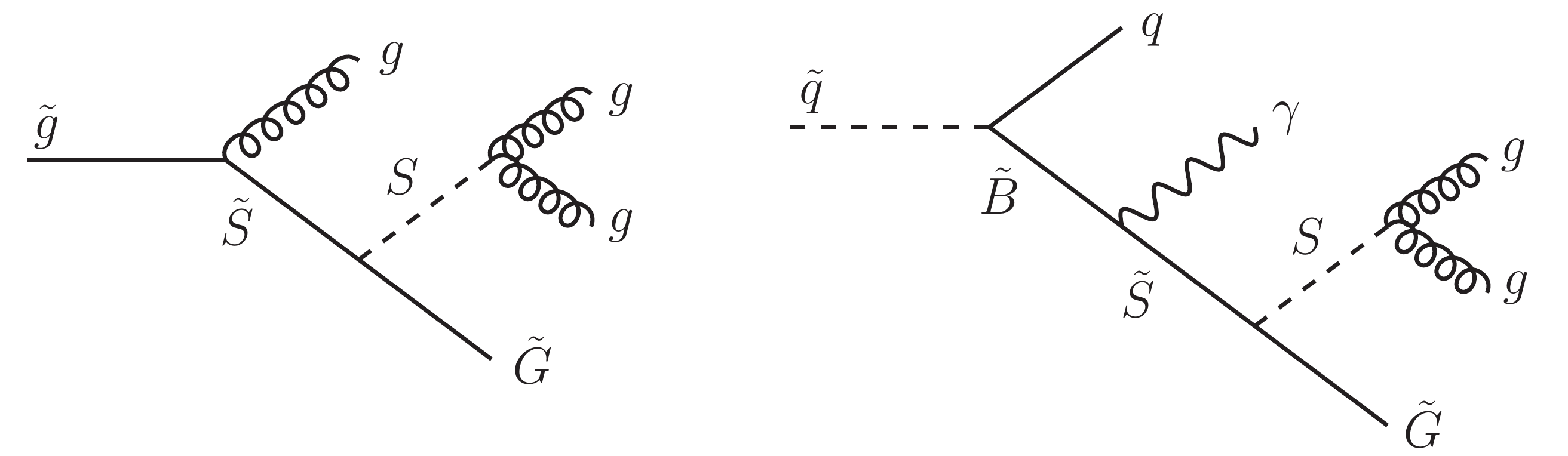}
\caption{In this section, we will consider the LHC limits on the two stealth diagrams shown in this figure.  The left diagram is an example of fully hadronic stealth SUSY, where the LOSP decays to jets and soft gravitinos.  The right diagram is an example where one photon is produced in each SUSY cascade.}
\label{fig:feynman}
\end{figure}

In this section we discuss several aspects of stealth LHC phenomenology.  As we saw above, the stealth SUSY framework can lead to many different final states, depending on the portal that connects the MSSM to the stealth sector, and  on the identity of the LOSP\@.  The common features of stealth phenomenology are low missing energy and high multiplicity final states, because the LOSP decays through the stealth sector to SM particles.  We begin by evaluating the limits that the LHC can now set on stealth SUSY\@.  In section~\ref{sec:HadronicLimit}, we discuss the limit on gluinos that decay to a fully hadronic final state, as in the left of figure~\ref{fig:feynman}.  In section~\ref{sec:GammaLimit}, we evaluate the limit on stealth decays that produce photons and jets as in the right of figure~\ref{fig:feynman}.  Finally, in section~\ref{sec:DisplaceMet}, we show that displaced vertices can lead to extra missing energy, although the size of the effect is typically small.

\subsection{Limit on Stealth with Jets and MET} \label{sec:HadronicLimit}
Here we present the limit on fully hadronic stealth decays.  We focus on gluino pair production, where the gluinos decay through the process shown to the right of figure~\ref{fig:feynman}.  Each gluino decays to three jets and a gravitino.  This is the dominant final state in the $S Y \bar Y$ model with a gluino LOSP\@.   The momentum of the gravitino, and therefore the amount of missing energy, depends on the mass splitting, $\delta m$ between $S$ and $\tilde S$.  We consider the limit coming from jets plus missing energy, which will allow us to evaluate how small $\delta m$ must be in order to sufficiently suppress missing energy and hide SUSY\@.

\begin{table}[h!]
\begin{center}
\begin{tabular}{|c|c|c|c|}
\hline
search & luminosity fb$^{-1}$  & $\delta m$ limit  [GeV] & ref.\\
\hline \hline
CMS jets + MET & $36 \times 10^{-3}$ & 17 & \cite{Collaboration:2011ida} \\
CMS jets + MET & 1.1 & 12 & \cite{CMS-PAS-SUS-11-004} \\
ATLAS 6-8 jets & 1.34 & 13 & \cite{Aad:2011qa} \\
ATLAS 2-4 jets & 1.04 & 35 & \cite{Aad:2011ib} \\
CMS $\alpha_T$ & 1.14 & 38 & \cite{Collaboration:2011zy} \\
CMS $M_{T2}$ & 1.1 & 45 & \cite{CMS-PAS-SUS-11-005} \\
\hline
\end{tabular}
\end{center}
\caption{\label{tab:JetMetSearches}
The LHC searches for jets and missing energy included in our analysis.   The ``$\delta m$ limit" column lists the value of the splitting in the stealth sector below which the search sets no limit at all, for a 100 GeV singlino.  We have included all of the jets plus missing energy searches conducted with 1~fb$^{-1}$, except for the CMS search based on the razor variable~\cite{CMS-PAS-SUS-11-008}, where we expect similar results.  For comparison we have also included the 2010 search that we found to set the strongest limit on stealth SUSY~\cite{Stealth}.
}
\end{table}

In our previous paper~\cite{Stealth}, we considered the limit on the same process coming from the 2010 data, $L \sim 35~\unit{pb}^{-1}$.   We found that the strongest limit in 2010 comes from the high $H_T$ channel of the CMS jets plus missing energy search~\cite{Collaboration:2011ida}.  With this search, we found no limit when $\delta m < 17$~GeV, fixing the singlino mass to 100 GeV and restricting to gluinos heavier than 200 GeV\@.  Here, we consider the limit that is set by various 2011 jets plus missing energy searches, all with about 1~fb$^{-1}$, listed in table~\ref{tab:JetMetSearches}, that were carried out by ATLAS and CMS\@.  We use pythia~6.4~\cite{pythia} for event generation, PGS~4~\cite{pgs} as a crude detector simulator, and $K$~factors for gluino production from Prospino~\cite{prospino}.  The limit that we find is shown in figure~\ref{fig:stealthlimits}, as a function of the gluino mass and $\delta m$, for a singlino mass of 100 GeV\@.  We find no limit for $\delta m < 12$~GeV, and this comprises the stealth regime where SUSY is hidden from searches that demand missing energy.  In the large $\delta m$ regime, the gluino mass limit extends to about 800~GeV\@.  In 2011, as in 2010, we find that the strongest limit on stealth comes from the CMS high $H_T$ search strategy, which is not surprising because this search demands less missing energy than the other searches, compensating this with a tighter cut on $H_T$.

\FIGURE[h]{
\includegraphics[scale=0.55]{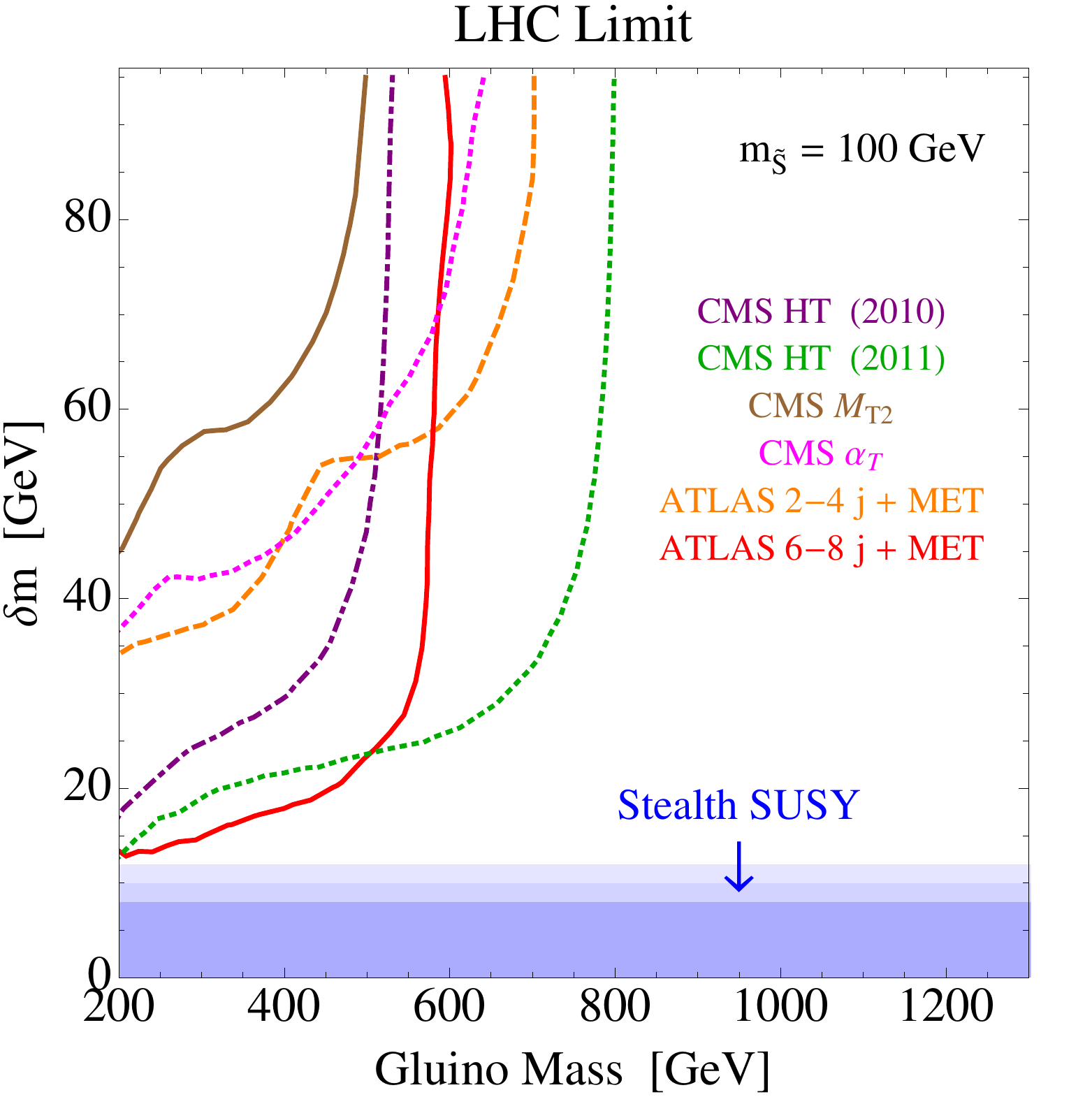} 
\caption{
The limits from searches for jets plus missing energy on stealth SUSY with fully hadronic decays, shown as a function of the gluino mass and the mass splitting in the stealth multiplet, $\delta m$.  We find that hadronic stealth SUSY is completely unconstrained by these searches when $\delta m \lesssim 10$~GeV\@.  Interestingly, the limit coming from 2011 searches is comparable to the limit coming from 2010 searches in the small $\delta m$ regime.
}
 \label{fig:stealthlimits}
}

One surprising feature of the limits shown in figure~\ref{fig:stealthlimits} is that, in the small $\delta m$ limit, the 2011 searches present only a modest improvement over the 2010 limit, despite the factor of 30 increase in luminosity.  The reason for this is that the 2011 searches have tighter cuts and therefore significantly lower acceptance than the 2010 searches.     The acceptance for stealth SUSY to pass these cuts is shown in figure~\ref{fig:stealthaccept}, where we compare the acceptance of the CMS high $H_T$ searches from 2010 and 2011.  The 2010 version of the search demanded missing energy above 150~GeV and $H_T>500$~GeV\@.  The 2011 version increased the missing energy cut to 200~GeV and the $H_T$ cut to 800~GeV\@. 

\FIGURE[h]{
\includegraphics[width=\textwidth]{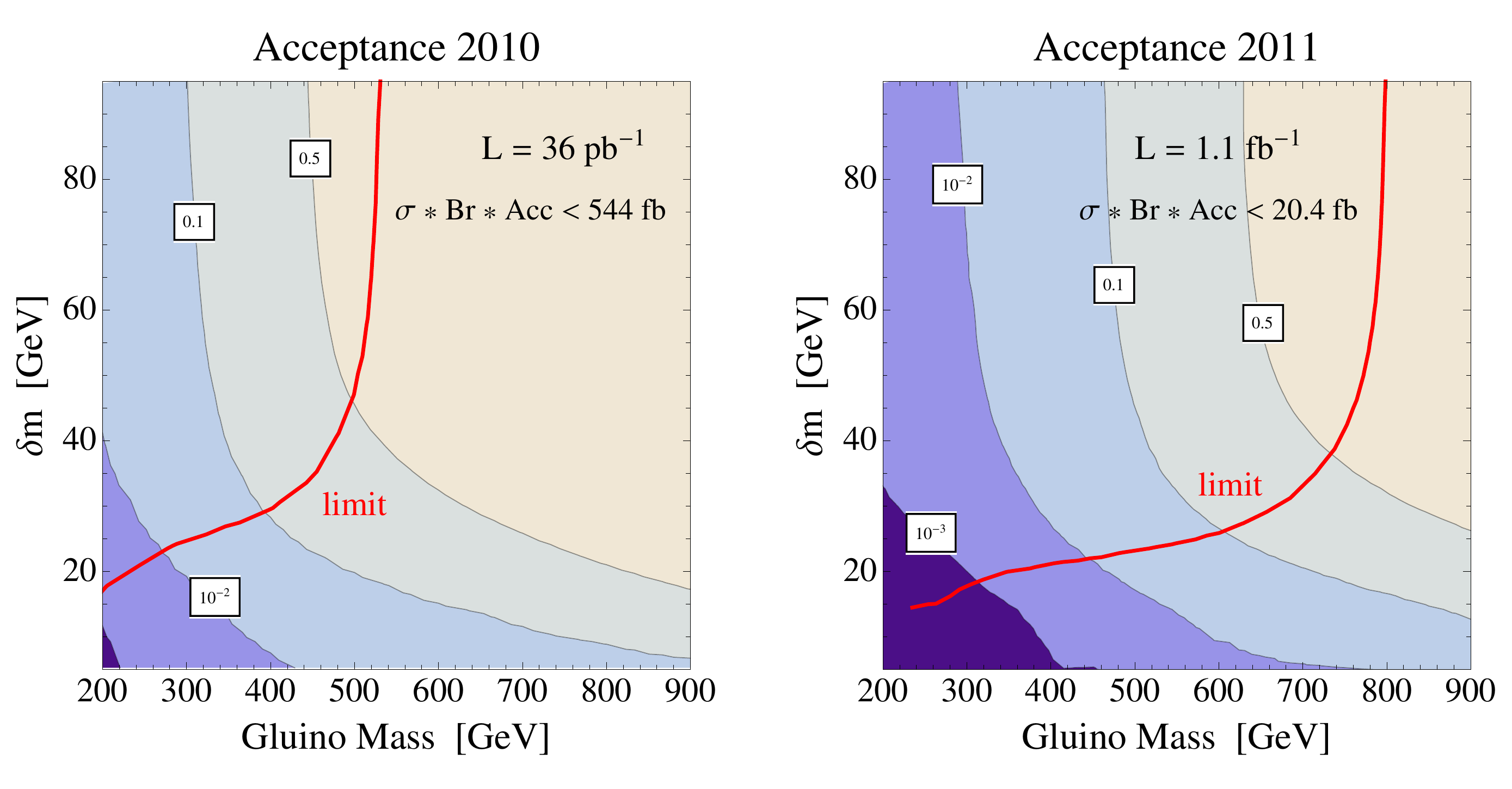} 
\caption{The acceptance for stealth SUSY events to pass the high $H_T$ selection of the 2010 and 2011 CMS searches for jets and missing energy.  These are the two searches that dominate the 2010 and 2011 limits, respectively.  Despite the large increase in luminosity in 2011, we find competitive limits in the small $\delta m$ regime because the increase in data is compensated by the decrease in acceptance that results from the tighter 2011 cuts.}
\label{fig:stealthaccept}
} 

\subsection{Limit on Stealth with Photons}
\label{sec:GammaLimit}

Now we consider the LHC limit on stealth decays that produce photons.  We focus on squark pair production, where each squark decays to 1 photon and 3 jets, as to the right of figure~\ref{fig:feynman}.  Every event contains two hard photons whose presence can be used to greatly reduce the QCD background.  We find that a limit can already be set on this scenario using the $\gamma \gamma$ mass distribution, which has been used by CMS~\cite{Chatrchyan:2011fq} and ATLAS~\cite{ATLAS:2011ab} to set limits on KK gravitons with about 2~fb$^{-1}$.  Although stealth SUSY does not produce a narrow $\gamma \gamma$ resonance, colored production can lead to a large enough cross-section to exceed the measured spectrum.

In order to check the LHC limit, we consider a spectrum where all light flavor squarks are degenerate and decay according to the diagram of figure~\ref{fig:feynman}.  We fix the bino mass to 300~GeV, the singlino mass to 100~GeV, and the stealth scalar mass to 95~GeV\@.  The right side of figure~\ref{fig:gamma} shows the $m_{\gamma \gamma}$ spectrum measured by CMS compared to the stealth spectrum that would be produced by $m_{\tilde q} = 500$ and 700~GeV\@.  We use Pythia~6.4~\cite{pythia} for event generation and we simulate the photon isolation ourselves from the hadronized Pythia output.  We use Prospino~\cite{prospino} for the NLO squark production cross-section.  To the right of figure~\ref{fig:gamma} we show the 95\% C.L. limit that we derive on the degenerate squark mass, $m_{\tilde q} \gtrsim 700$~GeV\@.  In order to derive this limit, we have binned the spectrum in 100~GeV bins, and we require that no bin exceed the 95\% C.L. limit using the CL$_s$ statistic~\cite{Junk:1999kv}.

To summarize, we have found that stealth decays to photons are not too stealthy, and $m_{\tilde q} \gtrsim 700$~GeV is required for degenerate light flavor squarks.  However, this limit is weaker than the limit on degenerate squarks in regular SUSY from jets plus missing energy, $m_{\tilde q} \gtrsim 900$~GeV, after 1~fb$^{-1}$.  We believe that there is still reach for a dedicated analysis to discover stealth decays to photons because the background can be further reduced by demanding a large $H_T$ of jets in addition to the presence of two hard photons.  Furthermore, it may be possible to reconstruct the $\gamma j j$ resonance, as we discussed in our previous paper~\cite{Stealth}.

We also note that certain stealth models include narrow, low mass, $\gamma \gamma$ resonances, and the branching ratio into these resonances can be used to constrain the total SUSY cross-section.  For example, the vectorlike portal of section~\ref{subsec:singlet} leads to a small branching ratio of the stealth singlet to $\gamma \gamma$. In the benchmark model of this type from our previous paper~\cite{Stealth}, the singlet branching ratio to photons was $4 \times 10^{-3}$.  A branching ratio of this size is constrained by the $5\unit{fb}^{-1}$ higgs searches~\cite{CMS-PAS-HIG-11-030, ATLAS-CONF-2011-161} if the singlet mass is within the search region of $m_s \approx 100-150$~GeV.  The limit on $\sigma \times \mathrm{Br}$ is  on the order of 0.1~pb, which constrains the SUSY cross-section to be smaller than about 50~pb.  This requires the gluino to be heavier than about 300 GeV.

\FIGURE[h]{
\includegraphics[width=\textwidth]{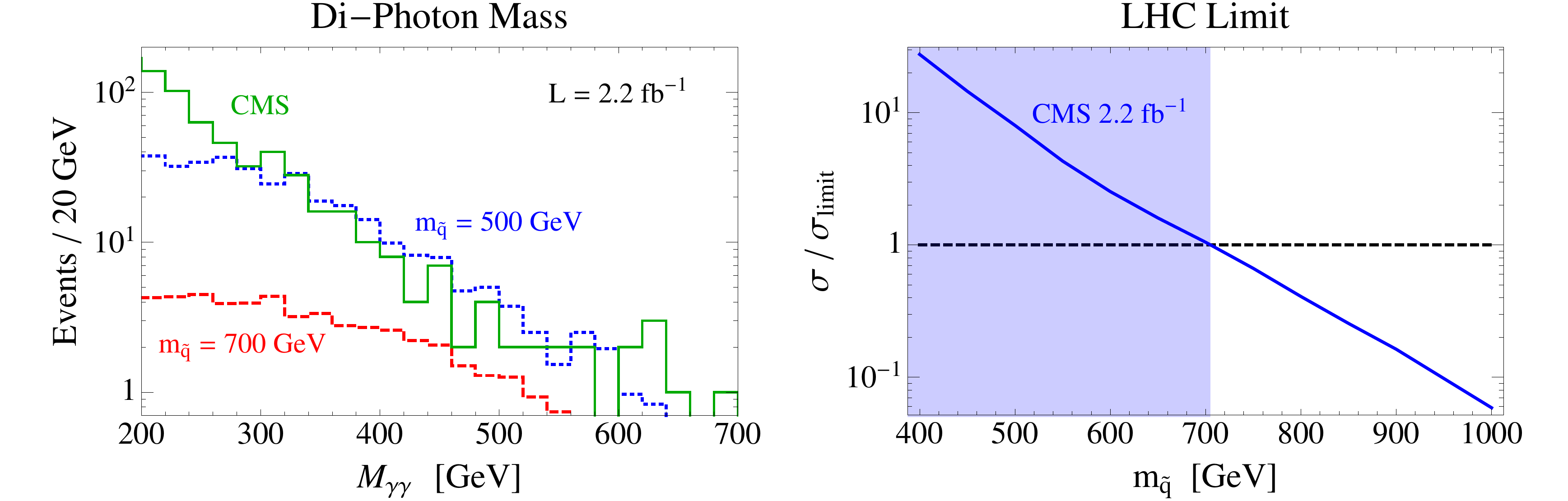} 
\caption{Stealth SUSY decays that produce photons, $\tilde N_1 \rightarrow \gamma j j$, are constrained by the $\gamma \gamma$ mass distribution.  The left plot shows $m_{\gamma \gamma}$ measured by CMS~\cite{Chatrchyan:2011fq} with 2.2~fb$^{-1}$ and the mass distribution that would be produced by degenerate light flavor squarks that decay to a neutralino that decays to $\gamma jj$.  We have fixed the neutralino mass to 300~GeV, the singlino mass to 100~GeV, and the stealth multiplet splitting to 10~GeV\@.  The right plot shows the ratio of the signal cross-section to the 95\%~C.L. limit that can be derived from the $\gamma \gamma$ spectrum.  We find that $m_{\tilde q} \lesssim 700$~GeV is excluded.}
\label{fig:gamma}
} 

\subsection{Missing $E_T$ at Long Lifetimes}
\label{sec:DisplaceMet}

\begin{figure}
\hbox{\hspace{5cm}\includegraphics[scale=0.4]{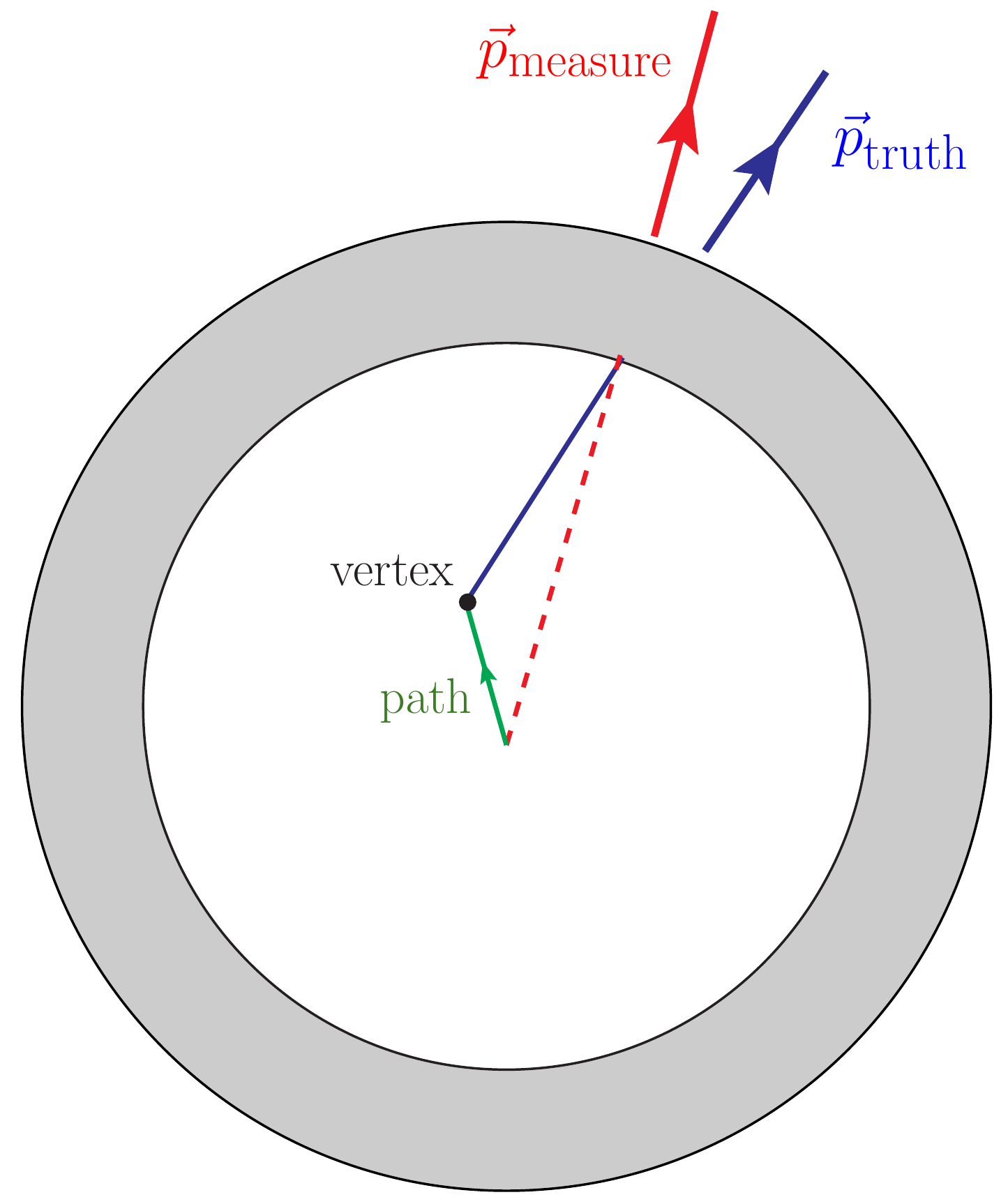} }
\caption{Jets that originate from a displaced vertex will be measured to have the incorrect momentum because the momentum of a jet is normally determined by pointing the calorimeter hits back to the primary vertex.   This momentum scrambling can lead to missing energy in stealth SUSY events with displaced vertices.}
\label{fig:displace_diagram}
\end{figure}

Here we wish to present an effect mentioned in~\cite{Stealth} in more detail. If missing transverse energy is computed from energies in the calorimeter, the pseudorapidity and azimuthal angle $(\phi,\eta)$ may be used as detector coordinates relative to the primary vertex to compute the momentum of the particle that deposited the energy, assuming it originated at the collision point. This will continue to give the correct momentum for particles originating at displaced vertices, as long as the decay products continue along the line of flight of the decaying particle. More generally, however, it will lead to a mismeasurement of momentum, as shown schematically in figure~\ref{fig:displace_diagram}.

\FIGURE[h]{
\includegraphics[width=\textwidth]{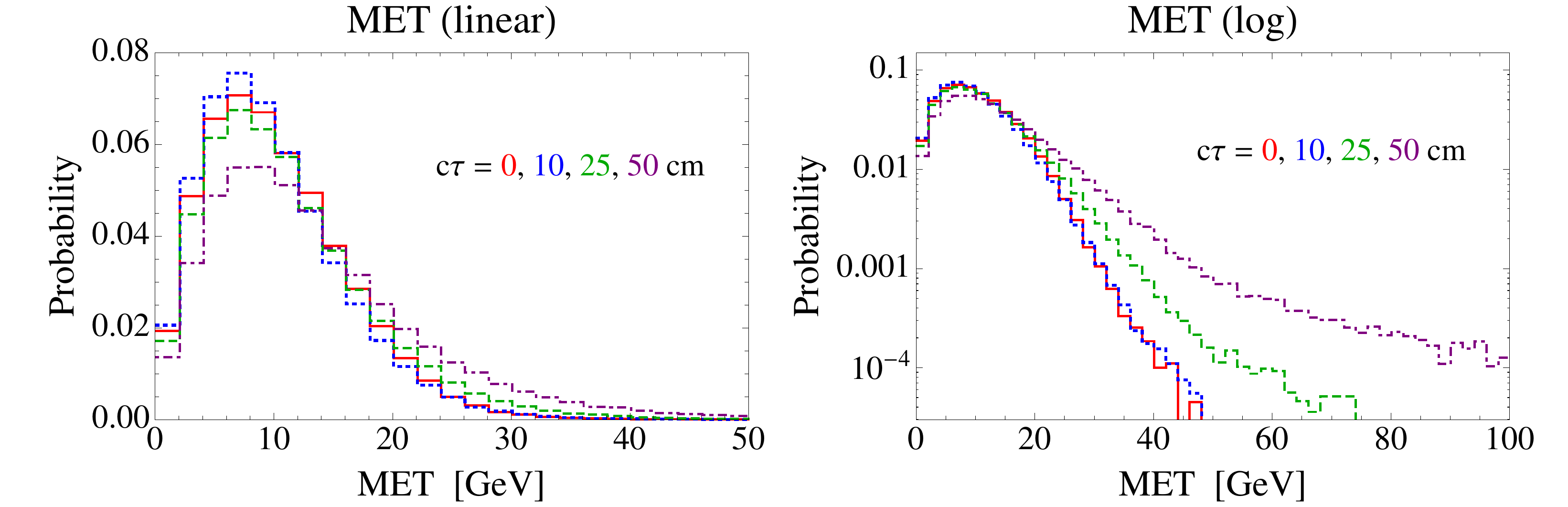}
\caption{The missing $E_T$ distribution for different singlino lifetimes, shown separately on linear and logarithmic scales.  We find that displacement does enhance the missing energy tail although the effect is mild, even for displacements of 10s of cm.  Here we have fixed the singlino mass to 100 GeV and the stealth multiplet splitting to 10 GeV\@.}
\label{fig:displacemet}
}

\begin{figure}[h]
\begin{center} \includegraphics[width=0.57\textwidth]{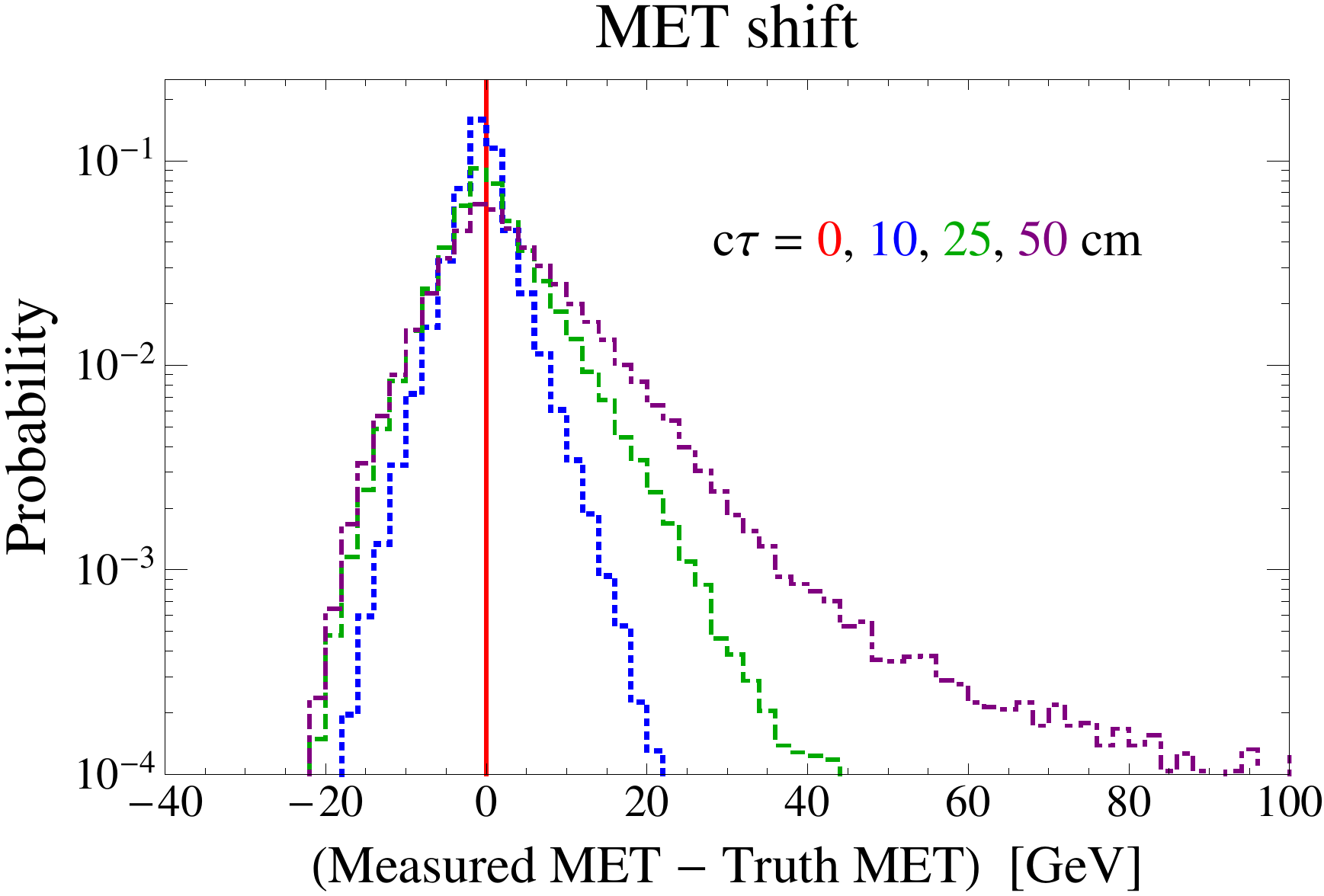} \end{center}
\caption{The difference of the measured missing energy and the Monte Carlo truth missing energy, for difference singlino lifetimes.  We see that larger lifetimes lead to a larger difference, and that the missing energy is more likely to be increased than decreased.}
\label{fig:displacemetdiff}
\end{figure}

To illustrate this effect, we have simulated 300 GeV gluinos decaying to 100 GeV singlinos in the decay chain ${\tilde g} \to g {\tilde S},~{\tilde S} \to S {\tilde G},~S \to gg$, shown to the left of figure~\ref{fig:feynman}. We compute the distribution of decay locations for the ${\tilde S} \to S {\tilde G}$ step, for various singlino lifetimes $c\tau$. Assuming a calorimeter located 1.25 meters from the beamline, we then compute the measured missing $E_T$ based on the locations where gluons from the $S$ decay will hit the calorimeter.\footnote{A more complete treatment, which we leave for further work, would also need to include the calorimeter shower shape, which can change the jet 4-vector for very large displacements.}  If the scalar $S$ is not extremely boosted, the $S \to gg$ decay has a wide opening angle, and the location of the calorimeter energy can vary significantly as the lifetime is varied. In Figure~\ref{fig:displacemet}, we show the resulting missing $E_T$ distribution for various $c\tau$ values on both linear and log scales, in the case of a splitting $m_{\tilde S} - m_S = 5$ GeV\@. The distributions peak in about the same place at all lifetimes, but the high missing $E_T$ tails become much larger at lifetimes of order the detector radius. (Only those events for which the decay happens before the calorimeter radius are plotted.) For lifetimes of 10 cm, though, the distribution has hardly changed at all. This is reassuring and in accord with our geometric intuition.    A similar illustration appears in Figure~\ref{fig:displacemetdiff}, this time illustrating the difference between the measured missing $E_T$ and the true value.  Since the overall missing energy distribution is not too sensitive to this effect, the limit we found in section~\ref{sec:HadronicLimit} carries over to the case where the singlino has moderate lifetime.

Of course, if the tracker is used instead of just the calorimeter, events with 10 centimeter lifetimes will look dramatically different from those with only prompt decays. We encourage searches for macroscopically displaced vertices or anomalous jets without tracks pointing at the calorimeter deposits, with all possible trigger paths.

\section{Discussion}
\label{sec:discussion}
\setcounter{equation}{0}
\setcounter{footnote}{0}

We are now truly in the LHC era, with 5/fb of data accumulated by each experiment and numerous  beyond the SM analyses completed using 1 to 2/fb.   Already, the landscape of possible theories has changed dramatically. Gluinos, in most realizations of $R$-parity preserving SUSY, are excluded up to masses of 700 GeV or more. Squarks of the first and second generation are similarly constrained, pointing to either flavored mediation of SUSY breaking or tuning in the MSSM\@.
The possible Higgs discovery at close to 125 GeV also points either to extensions of the MSSM or to fine-tuning in SUSY.  We must be prepared to move beyond the MSSM if we are to continue to take SUSY seriously as a natural theory of the electroweak scale.

In light of the emerging LHC data, it is becoming increasingly important to consider extensions of the MSSM that are less constrained by the LHC\@.  By surveying such extensions, new collider searches can be motivated in order to fill the gaps in existing search strategies and lead the way to unexpected discoveries.  We have presented a rich class of models that realize this idea through the stealth SUSY mechanism.  Missing energy is removed, and in its place are high multiplicity final states with rich resonance structures and the possibility of observably displaced vertices.  If the end of the 7 (or possible 8) TeV run at the LHC fails to bring a SUSY discovery in the missing energy tails, then it will become essential to assess the experimental status of stealth SUSY.  There remain many promising and unconstrained channels where natural SUSY may reveal itself.

\section*{Acknowledgments}
We thank Nima Arkani-Hamed, Cliff Cheung, Tim Cohen, Michele Papucci, Daniel Phalen, and Aaron Pierce for helpful conversations. MR would like to thank the KITP for its hospitality while part of this work was being completed (supported in part by the National Science Foundation under Grant No.~PHY05-51164). JF is supported by the DOE grant DE-FG02-91ER40671. MR is supported by the Fundamental Laws Initiative of the Center for the Fundamental Laws of Nature, Harvard University.  J.T.R. is supported by a fellowship from the Miller Institute for Basic Research in Science. 

\appendix

\section{Details of the High-Scale $SY{\bar Y}$ Model}
\label{app:viable}

We will consider an explicit model that can work with high-scale SUSY breaking, as briefly outlined in Section~\ref{subsec:viable}. We have argued in Section~\ref{subsec:tadpole} that, to avoid tadpoles that spoil the stealth mechanism, we wish to charge our field $S$ under a symmetry. To keep the same spirit as the models of~\cite{Stealth}, we will take this to be a discrete gauge (nonanomalous) symmetry. Furthermore, as we argued in Section~\ref{subsec:stealthbreaking}, a simple mass term of the form $mS^2$ or $m S {\bar S}$ in conjunction with high-scale breaking leads to a $B$-term soft mass that is not compatible with stealth phenomenology. Thus, we would like to write down a model with only dimension-3 operators in the superpotential, which nonetheless dynamically generates a mass for $S$ and spontaneously breaks the discrete gauge symmetry at a scale low enough that the induced tadpole is safely small.

Suppose we introduce a new SU(4) gauge group with 3 flavors $Q_i$, as well as 3 new vectorlike fields in the ${\bf 5}+{\bf \bar 5}$ of the SM SU(5), $Y_i, {\bar Y}_i$. We also introduce three singlets $S_i$. We consider a superpotential:
\beq
W = \sum_{i=1}^3 S_i \left(y Q_i {\bar Q}_i + \lambda Y_i {\bar Y}_i\right) - \frac{1}{3} \kappa \sum_{i = 1}^{3} S_i^3.
\eeq
This superpotential respects an anomaly-free ${\mathbb Z}/3$ discrete symmetry under which all fields have charge 1 (see for instance~\cite{Csaki:1997aw} for a guide to discrete anomalies). The choice of $W$ breaks the flavor symmetry acting on the $Q$'s and $Y$'s down to a product of U(1)s. This is an admittedly ugly tactic to avoid having a moduli space of supersymmetric vacua, because the massless modes on the moduli space can be problematic for us. This potential will give $S$ a VEV and hence a dynamical mass, as in~\cite{YanagidaMu,Dine:2009swa}. The SU(4) gauge group confines and generates an ADS superpotential~\cite{Affleck:1983rr},
\beq
W_{\rm eff} = \frac{\Lambda^9}{\det\left({\bar Q}Q\right)}.
\eeq
In the presence of this dynamical mass scale $\Lambda$, the theory has a supersymmetric vacuum
\beqs
\left<S_i\right> & = & \left(\frac{y^3}{\kappa^4}\right)^{1/9} \Lambda \label{eq:vevS} \\
\left<Q_i {\bar Q}_j\right> & = & \left(\frac{\kappa}{y^3}\right)^{1/9} \Lambda^2 \delta_{ij}. \label{eq:vevQ}
\eeqs
In the low-energy theory, instead of working in terms of the $Q, {\bar Q}$ fields, we will work in terms of the meson fields $M_{ij}$. In particular, we will assume that a canonically normalized meson is related to the quark fields by a matching condition:
\beq
Q_i {\bar Q}_j = \nu \Lambda M_{ij},
\eeq
with $\nu$ an order-one number. Let us compute the mass spectrum; it is easiest to work with the fermions. $W_{\rm eff}$ gives rise to Dirac masses $\nu^2 y^{5/3} \kappa^{-5/9} \Lambda \psi_{M_{ij}} \psi_{M_{ji}}$ for the off-diagonal mesinos, while the diagonal $M$ fields mix into two states with that mass and one state (Tr $M$) with four times that mass. The singlinos get masses $-\kappa^{5/9} y^{3/9} \Lambda$ from the $S^3$ terms, and also mix with the diagonal mesinos through the $y\nu S_i M_{ii}$ terms. Our goal for this sector is to generate a supersymmetric mass for the singlet fields $S$, as well as for the $Y$ fields, without substantially affecting SUSY breaking for the $S$ fields. 

To calculate SUSY-breaking effects, we assume that the MSSM fields together with the $S, Y$, and $Q$ fields are all sequestered, and that AMSB is the dominant contribution to the soft masses~\cite{Randall:1998uk,GLMR}. The SU(4) confining sector poses a complication for computing soft masses, as do the $Y$ fields, as nonsupersymmetric thresholds can take AMSB off its usual RG trajectory~\cite{Pomarol:1999ie,Katz:1999uw,Sanford:2010hc}. If the new states are below the gravitino mass, one can evaluate the soft terms above their mass scale and then run down, taking the thresholds into account. If they are heavier than the gravitino mass, they will contribute non-AMSB scalar mass-squareds of order $\frac{m_{3/2}^4}{16\pi^2 M^2}$~\cite{Katz:1999uw}. We will take $\Lambda \gg m_{3/2}$ so that this is the case for the composite states. There are two interesting hierarchies of couplings to consider in which the $S_i$ and $M_{ij}$ states are not highly mixed. If $\kappa \ll y \ll 1$, the mesinos are heavy relative to the singlinos, and $\left<S\right> \gg \left<M\right>$. Because the $Y{\bar Y}$ mass is $\lambda \left<S\right>$, this implies very small values of $\lambda$ if the $Y$ fields are to be light enough for their interactions to contribute any sizable AMSB soft terms. As a result, even the suppressed $m_{3/2}^4/\Lambda^2$ contributions from the composite sector can be important, and the SUSY-breaking spectrum is not fully calculable.

The limit $y \ll \kappa \ll 1$ is the more viable one, and leads to mesinos much lighter than the singlinos. Let us make some numerical estimates in this case. We fix $m_{3/2} = 30$ TeV and $\Lambda = 125$ TeV, taking $y = 10^{-5},~\kappa = 10^{-3}$. (We set $\nu = 1$, for the unknown matching of canonically normalized mesons with fundamental quarks; other values could be accommodated by changing other parameters to absorb the difference.) Then the supersymmetric mass of the singlet states is 116 GeV, two of the meson states have supersymmetric masses of 40 MeV, and the other meson state has a supersymmetric mass of 120 MeV\@. The mixings are small, $\epsilon_{MS} \simlt 10^{-2}$. The singlet VEV is $\left<S\right> \approx 58$ TeV\@. We take the $Y$ fields to be below the gravitino mass, $m_Y = \lambda \left<S\right> = 3$ TeV, by choosing $\lambda = 0.05$. Because $\lambda \gg \kappa, y$, the dominant contribution to the AMSB soft term for the singlets comes from $\lambda$ and is given by:
\beq
{\tilde m}_S^2 = \frac{1}{2} \left|m_{3/2}\right|^2 \frac{d}{dt} \gamma^S_S = \frac{\left|m_{3/2}\right|^2}{\left(16\pi^2\right)^2} \left(35 \left|\lambda\right|^4 - \frac{80}{3} \left|\lambda\right|^2 g_3^2\right) \approx -2400~{\rm GeV}^2.
\eeq
Ignoring all other SUSY-breaking effects, this would correspond to a stealth splitting $m_{\tilde S} - m_S \approx 10$ GeV, precisely as desired. (Strictly speaking, this calculated soft mass is valid above the scale $m_Y$, though we expect running below $m_Y$ to make little difference.) There are a number of other sources of SUSY-breaking effects to check, however. Given the term $-\frac{1}{3}\kappa S^3$ in the superpotential, AMSB gives rise to a scalar trilinear ($A$-term) $\frac{1}{3} m_{3/2} \beta_\kappa S^3$, which in the presence of an $S$ VEV becomes a contribution to the scalar mass. However, $\beta_\kappa \propto \kappa \lambda^2/(16 \pi^2)$, and this contribution is negligible.

The next concern is the contribution from the meson states, which mix with the singlets. The mesons feel SUSY breaking in two ways: first, there are AMSB contributions from the superpotential term $W_{\rm eff} = \nu^{-3} \Lambda^6/\det M$; second, there are power-suppressed effects from the nonsupersymmetric thresholds at the scale $\Lambda$ associated with composite states that interact strongly with the mesons. Let's first tackle the AMSB contributions. Expanding around the VEV, $M = \left<M\right>+\delta M$, we have schematically $\Lambda^6/\det M\sim \left(\Lambda/\left<M\right>\right)^6 \delta M^3$. Numerically, for our parameters $\left<M\right> \approx 22 \Lambda$ and so the ADS superpotential, for the purpose of estimating AMSB soft terms, amounts to a Yukawa coupling of order $10^{-8}$ and makes no significant contribution. More significantly, we expect soft masses of order ${\tilde m}_M^2 \sim \frac{m_{3/2}^4}{16\pi^2 \Lambda^2}$ from interactions with the composite states. These should lift the scalar meson states to masses $\sim 500$ GeV\@. However, we expect that the $S$ scalars are insulated from these effects, as they do not interact directly with the composite states. Through mixing to a meson and back, we estimate a contribution $\delta{\tilde m}_S^2 \sim \epsilon_{MS}^2 \frac{m_{3/2}^4}{16\pi^2 \Lambda^2} \sim \left(0.3~{\rm GeV}\right) m_S$, negligible relative to the AMSB soft term $S$ acquired through its interactions with $Y,{\bar Y}$.

Finally, we observe that this model has a built-in stealth decay of a singlino to its scalar partner and a mesino, ${\tilde S} \to S {\tilde M}$, arising from the $S^3$ vertex and the small mixing $\epsilon_{MS}$. The non-stealthy decay ${\tilde S} \to M {\tilde M}$ is forbidden by additional mixing factors (smaller than the phase space suppression) and by the expectation that the scalar $M$ is lifted to larger masses by its interaction with the heavy composite states. The stealth decay ${\tilde S} \to S {\tilde M}$ has a lifetime of order 10 microns, and would appear prompt from the point of view of collider physics.

This gives a proof of principle that a stealth model compatible with high-scale SUSY breaking is possible. Sequestering and a discrete gauge symmetry enforce the basic structure of small SUSY-breaking splittings and absence of dangerous $B$-terms. However, we had to choose Yukawa couplings of order $10^{-3}$ and $10^{-5}$: not smaller than those we have seen in nature, but perhaps uncomfortably small. Also, due to confined, strongly interacting states at relatively low masses that affect the SUSY-breaking mass terms for mesons mixed with the stealth fields $S$, the model is not as calculable as we might prefer. Finally, the breaking of a discrete symmetry raises the specter of domain wall problems in cosmology, much as in the NMSSM, and a somewhat more sophisticated version of the model could be needed to evade them. Stealth SUSY, then, seems to find a slightly less congenial home in the context of high-scale SUSY breaking than it did for low-scale SUSY breaking. It would be interesting to explore further whether alternative models might work without such small Yukawa couplings and light composites.

\section{SUSY Breaking for the Baryon Portal with $S^2 N$}
\label{app:s2n}

We will now give details of AMSB for the case of the baryon portal, $Sudd$, along with an $S^2 N$ operator to allow decays to a baryon-charged invisible fermion. We will sequester both the MSSM sector and the stealth sector. In the low-energy theory below the $U$ or $D$ mass, the Yukawa coupling $\lambda S^2 N$ contributes a positive soft mass. However, this is not the full answer. The fields $U$ or $D$ may be lighter than the gravitino mass, in which case they give nonsupersymmetric thresholds that take AMSB off its usual RG trajectory~\cite{Pomarol:1999ie,Katz:1999uw,Sanford:2010hc}. In that case, we should still apply the AMSB formulas, but only above the scale $M$, and then run the results down from the nonsupersymmetric threshold according to the usual RGEs. Above the scale $M$, new Yukawas also contribute. Thus, we should evaluate the AMSB soft masses using all interactions in the first line of~\ref{eq:fullW}. The result is:
\beqs
{\tilde m}_N^2 & = & \frac{\left|m_{3/2}\right|^2}{(16\pi^2)^2} \left(20 \left|\lambda\right|^4 + 12 \left|\lambda\right|^2\left|a\right|^2 + 4 \left|\lambda\right|^2 \left|\alpha\right|^2\right) \\
{\tilde m}_S^2 & = & \frac{\left|m_{3/2}\right|^2}{(16\pi^2)^2} \left(40 \left|\lambda\right|^4 + 36 \left|\lambda\right|^2\left|a\right|^2 + 15 \left|a\right|^2 - 16 \left|a\right|^2 \left(g_3^2 + \frac{g'^2}{3}\right)\right. \nonumber \\
& & ~~~ \left.+12\left|\lambda\right|^2\left|\alpha\right|^2+6 \left|a\right|^2\left|\alpha\right|^2+2\left|\alpha\right|^4 + 16\pi^2 \left|\alpha\right|^2 \gamma^X_X\right) \\
{\tilde m}_{\bar S}^2 & = & \frac{\left|m_{3/2}\right|^2}{(16\pi^2)^2} \left(4\left|\alpha\right|^2\left|\lambda\right|^2 + 3\left|\alpha\right|^2\left|a\right|^2 + 2\left|\alpha\right|^4 + 16\pi^2 \left|\alpha\right|^2\gamma_X^X\right) \\
a_\alpha & = & -\frac{\alpha m_{3/2}}{16\pi^2} \left(4\left|\lambda\right|^2 + 3\left|a\right|^2 + 2 \left|\alpha\right|^2 + 16\pi^2 \gamma_X^X\right)
\eeqs
where $\left|a\right|^2$ implicitly means $\sum_i \left|a_i\right|^2$ and the anomalous dimension $\gamma^X_X$ cannot be calculated without a UV completion of the mechanism by which $\left<X\right>$ is determined. Because $X$ has a VEV, the trilinear $a_\alpha$ term becomes a $B$-term:
\beq
BS{\bar S} = -\frac{m_S m_{3/2}}{16\pi^2} \left(4\left|\lambda\right|^2 + 3\left|a\right|^2 + 2 \left|\alpha\right|^2 + 16\pi^2 \gamma_X^X\right) S{\bar S}.
\eeq
The $-g_3^2 \left|a\right|^2$ term in ${\tilde m}_S^2$ threatens to give a sizeable wrong-sign soft mass if we do not take $a \ll \lambda$. Furthermore, the $B$-term splits the eigenstates and tends to produce a scalar lighter than the fermion. This is difficult to avoid, as ${\tilde m}_S^2\gg \left|B\right|$ typically implies a large enough $\lambda$ that splittings are not stealthy. This differs from the vectorlike portal with AMSB discussed in Section~\ref{subsec:viable} because in that case we could take the piece of a soft mass proportional to $-g_3^2$ to dominate. Here, that would give the wrong sign.

Anomaly mediation can thus easily give stealth-size splittings, by taking the couplings to be small, but it tends to produce well-mixed scalar states, one of which is lighter than its fermionic partner. The outcome is a partially stealthy spectrum: about half the time, the LOSP will decay to the heavier scalar $S$, which will decay to its partner ${\tilde S}$ and a soft fermion $\tilde{N}$. The other half of the time, the decay will proceed to the lighter scalar $S$. Its decay is not stealthy, because it will go through an off-shell ${\tilde S}$ (in a four-body decay). Comparing to Equation~\ref{eq:suddsinglinolifetime}, we can estimate that for $\lambda,\lambda_{uds} \sim 0.1$, $m_{\tilde S} \approx 200$ GeV, and gluinos and squarks at 500 GeV, the lifetime would be of order a few centimeters. In these decays the $N$ is sharing energy with several visible particles, and for small splittings the singlino propagator is enhanced when the $N$ has minimal energy, competing with phase space suppression in that region. Thus, we expect that missing energy is strongly reduced even in the half of the decays that are not, strictly speaking, stealthy. 

Alternatively, we can simply appeal to an extension of AMSB to produce soft terms comparable in size that lift both scalar states above the fermion. Because minimal AMSB in the MSSM predicts tachyonic sleptons, a number of models have been proposed to generate additional contributions to SUSY-breaking for scalars of the same order as AMSB contributions without changing the basic paradigm of sequestering (e.g.~\cite{Pomarol:1999ie, Katz:1999uw, ArkaniHamed:2000xj}). Such models can be adapted to produce a fully-stealthy sequestered version of the $Sudd$ model with $S^2 N$ decay.

\end{document}